\documentclass[
english,
10pt,%
notitlepage,%
twocolumn,
nofootinbib,
superscriptaddress,%
nolongbibliography,
noshowpacs,%
centertags]%
{revtex4-1}

\usepackage{graphicx, caption2}
\usepackage{rotating, amssymb}

\newcommand{\Teff}{$T_{\mbox{eff}}$}
\newcommand{\lgg}{\rm{lg}\,$g$}
\newcommand{\Vt}{$V_t$} 
\newcommand{\Mo}{$M_{\odot}$}
\newcommand{\kms}{$km\,s^{-1}$}
\def\refitem#1{\relax}

\sloppypar

\begin{document}

\title{\bf Sodium in the Atmospheres of Thick-Disk Red Giants}

\author{\bf Yu. V. Pakhomov}
\affiliation{{\it Institute of Astronomy, Russian Academy of Sciences,
Pyatnitskaya ul. 48, Moscow, 109017 Russia}, pakhomov@inasan.ru}

\begin{abstract}
The parameters and elemental abundances of atmospheres for ten thick-disk red
giants was determined from high-resolution spectra by the method of model
stellar atmospheres. The results of a comparative analysis of the [Na/Fe]
abundances in the atmospheres of the investigated stars and thin disk red giants
are presented. Sodium in the atmospheres of thick-disk red giants is shown to
have no overabundances typical of thin-disk red giants.


\end{abstract}

\maketitle

\section*{INTRODUCTION}
\noindent
The abundances of chemical elements in the atmospheres of stars are known to
change in the course of their evolution. When investigating the chemical
composition of giants, the first to be detected were the CNO abundance
anomalies compared to the chemical composition of dwarf stars (see, e.g.,
\cite{1954AJ.....59R.192S, 1958ApJ...127..172G, 1959ApJ...129..700H, 
1968AJS....73Q..15G}). The observed
sodium overabundance in the atmosphere of the red giant $\epsilon~Vir$ was
first mentioned in 1963 \cite{1963ApJ...137..431C}. Subsequently, sodium
overabundances were detected in red giants of the Hyades open cluster
\cite{1964ApJS....9...81H}. In another paper \cite{1968ApJS...16....1H},
the same authors found several giants with a [Na/Fe] overabundance and may have
been the first to point out a systematic difference in sodium abundance
between red giants and dwarfs. They hypothesized that sodium might be
synthesized in some red giants. Slightly later, in 1970
\cite{1970A&A.....7..408C} noted that the [Na/Fe] abundance increases with
temperature parameter $\theta$=5040/\Teff. In other words, the [Na/Fe] abundance
in cooler stars turned out to be, on average, higher, while giants are in the
majority among these stars, i.e., the sodium abundance difference between
dwarfs and giants is also noticeable, although the authors did not reach this
conclusion. 

\cite{2001ARep...45..301B} were the first to describe the
dependence of the [Na/Fe] overabundances in the atmospheres of normal red giants
on surface gravity. This dependence turned out to be similar to that discovered
previously for supergiants \cite{1981BCrAO..63...68B, 1981BCrAO..64....1B}.
It was explained by the hypothesis that sodium could be produced in the
$^{22}$Ne(p,$\gamma$)$^{23}$Na reaction entering the neon─sodium cycle of
hydrogen burning in the cores of main sequence stars and could then be brought
from deep layers into the stellar atmosphere through
developing convection as the star evolves from the main sequence to the red
giant branch \cite{1983BCrAO..66..119B}. The calculations performed by
\cite{1987SvAL...13..214D, 1988SvAL...14..435D} confirmed the
validity of this hypothesis.Further studies revealed [Na/Fe] overabundances in
the atmospheres of various classes of red giants: moderate and classical barium
stars \cite{2002ARep...46..819B, 2004ARep...48..597A}, super-metal rich
stars \cite{2009ARep...53..685P}. A database of characteristics for red giants
was created from the accumulated material at the Institute of Astronomy, the
Russian Academy of Sciences. Our studies using the database revealed several
stars with a [Na/Fe] underabundance relative to the observed dependence on
surface gravity \cite{2005ARep...49..535A, 2009ARep...53..685P}. A
comparative analysis showed that these stars are distinguished by slightly
higher space velocities. To confirm this conclusion, we investigated the
chemical composition of red giants with high Galactic velocities
\cite{2012AstL...38..101P} and pointed out that 12 of the 14 investigated
stars also have a [Na/Fe] underabundance.

There exist quite a few works devoted to the abundance analysis of Galactic
thin- and thick-disk stars. These works are based on spectroscopic observations
of dwarfs, because their atmospheric elemental abundances undergo no changes as
they evolve on the main sequence. Besides, in comparison with giants, dwarfs
form a larger group and their continuum spectrum exhibits a smaller number of
spectral lines, which increases the accuracy of abundance determinations.
Therefore, there are comparatively few works devoted to thick-disk giants. There
are even fewer studies of sodium in thick-disk giants (see, e.g.,
\cite{2012MNRAS.422.1562S, 2010A&A...513A..35A}). In these papers, the
dependences of the [Na/Fe] abundance on mass and metallicity are discussed, but
such an important factor as the evolutionary status is overlooked. Since the
sodium abundances in stellar atmospheres do not appear instantaneously but
gradually increase since the first deep mixing, the evolutionary stage of the
star should be taken into
account. In this paper, we perform a comparative analysis of the [Na/Fe]
abundance in the atmospheres of thin and thick-disk red giants as a function of
their metallicity and surface gravity, which changes during the lifetime of a
star. Thus, investigating the changes in the atmospheric sodium abundance of red
giants is necessary for understanding the stellar evolution, while the detected
sodium underabundance in thick disk giants requires a further study.

\section*{OBSERVATIONS}

\subsection*{Selection of Stars}

\begin{table*}
\setcaptionmargin{0mm}
\renewcommand\arraystretch{0.7}
\onelinecaptionsfalse
\captionstyle{flushleft}
\caption{The list of investigated stars with the membership probabilities ($p$)
in the Galactic thin and thick disks}
\label{tab:list}
\begin{tabular}{|c|c|c|c|c|c|c|c|c|}
\hline
N&HD    &$\alpha_{2000}$&$\delta_{2000}$&$m_V$&SpType&\multicolumn{2}{c|}{$p$,
\%} \\
&& &&  & &thin& thick\\
&& h:m:s&$\circ:\prime:\prime\prime$&  &  &disk& disk\\ 
\hline
 1&    249 & 00 07 22.56332 &+26 27 02.1686 &7.33 &K1IV  &  0 & 99\\
 2&   6555 & 01 06 38.62642 &+23 13 57.4834 &7.95 &K0III-IV & 15 & 84\\
 3&  10057 & 01 38 19.91771 &+02 35 09.8395 &6.92 &K0    &  1 & 97\\
 4&  24758 & 03 59 17.62739 &+57 59 13.6255 &8.67 &K0III &  0 & 99\\
 5&  37171 & 05 37 04.38211 &+11 02 06.0293 &6.00 &K5III &  0 & 99\\
 6&  80966 & 09 23 50.43122 &+34 32 53.3986 &7.17 &K0    &  0 & 99\\
 7& 180682 & 19 15 43.69429 &+40 21 35.7242 &6.96 &K0    &  0 & 99\\
 8& 203344 & 21 21 04.39426 &+23 51 21.4872 &5.57 &K1III &  0 & 99\\
 9& 211683 & 22 18 37.60252 &+10 21 18.7937 &7.73 &K2    & 20 & 79\\
10& 212074 & 22 21 27.94691 &+14 53 49.6214 &7.64 &K1IV  &  3 & 96\\
\hline
\end{tabular}
\end{table*}

\begin{table*}
\setcaptionmargin{0mm}
\renewcommand\arraystretch{0.7}
\onelinecaptionsfalse
\captionstyle{flushleft}
\renewcommand{\tabcolsep}{2pt} 
\caption{Kinematic parameters of the investigated stars: the Galactic velocity
vector components $(U,V,W)$ and Galactic orbital elements}
\label{tab:orbit}
\begin{tabular}{|c|c|c|c|c|c|c|c|c|}
\hline
N  &  HD  &       $U$      &       $V$     &       $W$     & $R_{min}$ & $R_{max}$ & $Z_{max}$ & $e$ \\
   &      &       \kms     &       \kms    &        \kms   & \kms      & \kms  
  & \kms       & \\
\hline
 1 &249   &   45.4$\pm$3.1  & -55.5$\pm$ 5.2 &  -78.6$\pm$ 5.3 & 6.00$\pm$0.14 &  8.93$\pm$0.04 &  1.46$\pm$0.09 & 0.20$\pm$0.01  \\
 2 &6555  &  -25.0$\pm$4.0  & -12.1$\pm$ 3.0 &  -68.3$\pm$ 5.8 & 7.98$\pm$0.06 & 10.04$\pm$0.13 &  1.21$\pm$0.10 & 0.11$\pm$0.01  \\
 3 &10057 &    3.6$\pm$7.1  & -46.8$\pm$ 7.0 &   55.6$\pm$ 1.3 & 6.90$\pm$0.24 &  8.70$\pm$0.02 &  1.38$\pm$0.02 & 0.12$\pm$0.02  \\
 4 &24758 & -123.7$\pm$4.3  & -23.8$\pm$ 7.9 &  -84.2$\pm$13.8 & 5.98$\pm$0.12 & 14.64$\pm$0.38 &  2.36$\pm$0.41 & 0.42$\pm$0.01  \\
 5 &37171 & -113.1$\pm$0.8  & -19.0$\pm$10.4 &   65.2$\pm$ 9.5 & 6.15$\pm$0.15 & 14.42$\pm$0.42 &  2.18$\pm$0.25 & 0.40$\pm$0.00  \\
 6 &80966 &  -23.4$\pm$14.0 &-105.4$\pm$23.8 &   75.5$\pm$ 9.5 & 3.92$\pm$0.56 &  8.83$\pm$0.08 &  2.15$\pm$0.19 & 0.39$\pm$0.06  \\
 7 &180682&  -89.7$\pm$8.7  & -11.6$\pm$ 4.2 &   59.6$\pm$ 5.7 & 6.72$\pm$0.10 & 12.70$\pm$0.31 &  1.70$\pm$0.13 & 0.31$\pm$0.02  \\
 8 &203344&   61.5$\pm$1.0  &-102.4$\pm$ 0.6 &  -70.1$\pm$ 2.5 & 3.69$\pm$0.02 &  8.86$\pm$0.01 &  1.21$\pm$0.05 & 0.41$\pm$0.00  \\
 9 &211683&   -7.1$\pm$4.4  & -20.2$\pm$ 3.3 &   50.2$\pm$ 5.7 & 8.12$\pm$0.09 &  8.75$\pm$0.11 &  1.06$\pm$0.08 & 0.04$\pm$0.01  \\
10 &212074&   -6.2$\pm$3.0  &  -5.9$\pm$ 5.5 &   61.5$\pm$ 8.7 & 8.34$\pm$0.02 &  9.94$\pm$0.27 &  1.46$\pm$0.16 & 0.09$\pm$0.01  \\
\hline
\end{tabular}
\end{table*}

\noindent

We selected the objects for our observations from the Hipparcos catalogue
\cite{2007A&A...474..653V} by analyzing the Galactic velocities ($UVW$)
calculated using reduced Hipparcos parallaxes and CORAVEL radial velocities
\cite{1987A&AS...67..423M}. The selection criteria were the (B-V) color indices
and calculated surface gravities corresponding to red giants; the calculated
Galactic velocities exceeding those typical of thin-disk stars (34.5, 22.5,
18.0)~\kms \cite{2005A&A...430..165F}, with $W<50$~\kms.

The list of program stars is presented in Table~\ref{tab:list}, where their
ordinal numbers, HD numbers, coordinates, V magnitudes, and spectral types are
given. The last two columns in Table~\ref{tab:list} provide the membership
probabilities of the program stars in the Galactic thin and thick disks, whose
kinematic characteristics were taken from \cite{2005A&A...430..165F}. The
probabilities were calculated using formulas from \cite{2004A&A...418..551M}.
It can be seen from the table that all stars can be assigned to thick-disk
objects with a high probability.

Table~\ref{tab:orbit} presents kinematic characteristics of the investigated
stars. These include the Galactic velocity vector $(U,V,W)$ relative to the Sun
and Galactic orbital elements: the perigalactic distance $R_{min}$, the
apogalactic distance $R_{max}$, the maximum orbital distance from the Galactic
plane $Z_{max}$, the eccentricity $e$, and the inclination $i$. The distance to
the Galactic center was assumed to be 8.5~kpc, while the necessary correction of
the velocities for the solar motion, (+10.2, +14.9, +7.8)~\kms, was taken
from~\cite{2005A&A...430..165F}. We calculated the orbital elements through
numerical integration of the stellar motion by Everhart's 15th-order method
using a three-component model Galactic potential \cite{1991RMxAA..22..255A}.
The integration accuracy was controlled by the conservation of the necessary
integrals of motion. For example, in ten orbital revolutions, the typical
relative error was $\Delta{}h/h$\textless{}$10^{-13}$ in angular momentum and
$\Delta{}E/E$\textless{}$10^{-8}$ in total energy. The errors in the space
velocities $(\sigma U,\sigma V,\sigma W)$ were calculated from the errors in the
stellar proper motions, radial velocities, parallaxes and the errors in the
solar velocity components relative to the local standard of rest. We calculated
the errors in the Galactic orbital elements based on the model Galactic
gravitational potential using the probable errors in the stellar space
velocities. It can be seen from Table 2 that all stars have a maximum orbital
distance from the Galactic plane $Z_{max}>1000$~pc, which exceeds considerably
the characteristic scale height for thin-disk objects, 90-325~pc
(\cite{1983MNRAS.202.1025G}, \cite{1996A&A...305..125R},
\cite{2001ApJ...553..184C}). More than half of the investigated stars have
orbits with eccentricities larger than 0.2. About half of the stars recede to a
distance of more than 10~kpc from the Galactic center when moving in their
orbits.

\subsection*{Spectroscopic Observations}

\begin{table}
\setcaptionmargin{0mm}
\renewcommand\arraystretch{0.7}
\onelinecaptionsfalse
\captionstyle{flushleft}
\caption{Parameters of atmosphere of the investigated stars, their masses, and
interstellar extinctions}
\label{tab:param}
\begin{tabular}{|c|c|c|c|c|c|c|c|c|c|c|c|}
\hline
N  &  HD    &\Teff &\lgg &\Vt  &[Fe/H]&  Mass     &$A_V$\\
   &        & K    &     &\kms &      & $M_\odot$  & m   \\
\hline
 1 &    249 & 4850 &2.96 &1.17 &-0.15 & 1.5$\pm$0.2& 0   \\
 2 &   6555 & 4720 &3.00 &1.15 &-0.09 & 1.2$\pm$0.2&$<$0.1 \\
 3 &  10057 & 4130 &1.70 &1.35 &-0.30 & 1.7$\pm$0.3& 0.3 \\
 4 &  24758 & 4680 &2.75 &1.15 & 0.11 & 1.4$\pm$0.2& 0   \\
 5 &  37171 & 4000 &1.25 &1.35 &-0.55 & 1.1$\pm$0.2& 0.5 \\
 6 &  80966 & 4550 &1.80 &1.50 &-1.01 & 1.2$\pm$0.3&$<$0.1 \\
 7 & 180682 & 4330 &1.90 &1.40 &-0.43 & 1.1$\pm$0.2& 0.3 \\
 8 & 203344 & 4770 &2.75 &1.30 &-0.09 & 1.8$\pm$0.2& 0   \\
 9 & 211683 & 4450 &1.80 &1.35 &-0.13 & 1.9$\pm$0.3& 0.3 \\
10 & 212074 & 4700 &2.55 &1.30 & 0.05 & 2.3$\pm$0.3&$<$0.1 \\
\hline
\end{tabular}
\end{table}

\noindent

The spectroscopic observations of the selected stars were performed in 2010 with
a two-band echelle spectrograph attached to a 2.16-m telescope at the Xinglong
station of the National Astronomical Observatories of China (NAOC). The
spectrograph operated in the red-band mode. The detector was a 2048x2048 CCD
array on which 45 spectral orders were recorded. The in the range from 5500 to
9830~\AA spectrograph resolution was $R=40\,000$; the signal-to-noise ratio in
the spectra was $S/N>100$. The \emph{echelle} package of the \emph{MIDAS}
software system was used for the preliminary spectroscopic data reduction, the
search for and extraction of the spectral orders, the wavelength calibration
using the spectrum of a thorium-argon lamp, and the spectrum normalization.

The equivalent widths of the selected spectral lines were measured in the EW
code that I developed. The code is a set of modules written in Perl and C, the
access to which is organized via a graphical user interface. The automatic
equivalent width determination module is the main one in the code. It uses the
nonlinear Levenberg-Marquardt algorithm from the PDL package. For each
measured spectral line with wavelength $\lambda_0$, the part of the spectrum
$\lambda_0\pm10\lambda_0/R$ in which the spectral lines are searched for is cut
out. An iterative deconvolution method is applied for a better line detection;
it allows the blends to be separated. More or less significant lines can be
identified by varying the number of iterations. Each spectral line is fitted by
a Gaussian, which is a normal approximation for most of the lines with
equivalent widths $<100$~m\AA. The investigated part of the 
spectrum is fitted by the sum of Gaussians $F(\lambda)=\sum_{i=0}^k
h_i*e^{-(\lambda_i-\lambda_0)^2/\sigma_i^2}$, where $h_i$ and $\sigma_i$ are the
and width of the $i$ spectral line. We take the depth of the observed line as
the initial approximation for $h$ and $\sigma=\lambda_0/R$ for the width. The
module operation result is a set of parameters of the Gaussians that best fit
the observed spectrum. The parameters of the investigated spectral line are
written in a file. In addition, information about the quality of the fit and
the degree of line blending is written. During its operation, the code displays
a theoretical spectrum of atomic and molecular lines that provides an additional
possibility for the selection of lines.

\section*{DETERMINATION OF STELLAR ATMOSPHERE PARAMETERS}

\begin{sidewaystable*}
\renewcommand{\tabcolsep}{2pt}
\setcaptionmargin{0mm}
\renewcommand\arraystretch{0.7}
\onelinecaptionsfalse
\captionstyle{flushleft}
\caption{Abundances of chemical elements in atmospheres of the studied stars}
\label{tab:abund}
\begin{tabular}{|c|cc|cc|cc|cc|cc|cc|cc|cc|cc|cc|}
\hline
      &\multicolumn{2}{c|}{HD 249}&\multicolumn{2}{c|}{HD 6555}&\multicolumn{2}{c|}{HD 10057}&\multicolumn{2}{c|}{HD 24758}&\multicolumn{2}{c|}{HD 37171}&\multicolumn{2}{c|}{HD 80966}&\multicolumn{2}{c|}{HD 180682}&\multicolumn{2}{c|}{HD 203344}&\multicolumn{2}{c|}{HD 211683}&\multicolumn{2}{c|}{HD 212074}\\
\hline
     &  N  &   [X/H]        &  N  &   [X/H]        &  N  &   [X/H]        &  N  &   [X/H]        &  N  &   [X/H]        &  N  &   [X/H]        &  N  &   [X/H]        &  N  &   [X/H]        &  N  &   [X/H]        &  N  &   [X/H]    \\
\hline
 NaI &   2 &  -0.23$\pm$0.04&   2 &  -0.03$\pm$0.09&   2 &  -0.26$\pm$0.08&   2 &   0.33$\pm$0.07&   1 &  -0.66         &   2 &  -1.22$\pm$0.02&   2 &  -0.36$\pm$0.04&   2 &  -0.03$\pm$0.11&   2 &  -0.07$\pm$0.15&   2 &   0.04$\pm$0.07\\
 MgI &   2 &  -0.03$\pm$0.02&   2 &   0.17$\pm$0.04&   2 &  -0.06$\pm$0.04&   2 &   0.21$\pm$0.03&   2 &  -0.29$\pm$0.08&   2 &  -0.60$\pm$0.06&   2 &  -0.12$\pm$0.04&   2 &   0.16$\pm$0.01&   2 &   0.08$\pm$0.05&   2 &   0.26$\pm$0.05\\
 AlI &   2 &   0.06$\pm$0.02&   2 &   0.20$\pm$0.01&   2 &   0.05$\pm$0.09&   2 &   0.20$\pm$0.01&   2 &  -0.19$\pm$0.05&   2 &  -0.72$\pm$0.02&   2 &  -0.10$\pm$0.04&   2 &   0.18$\pm$0.05&   2 &   0.15$\pm$0.04&   2 &   0.20$\pm$0.05\\
 SiI &   9 &  -0.15$\pm$0.05&   9 &   0.02$\pm$0.07&   4 &  -0.23$\pm$0.07&  12 &   0.18$\pm$0.08&   5 &  -0.48$\pm$0.07&   9 &  -0.65$\pm$0.03&   7 &  -0.25$\pm$0.04&   9 &   0.06$\pm$0.06&   8 &  -0.12$\pm$0.09&  11 &   0.08$\pm$0.08\\
 CaI &   4 &   0.03$\pm$0.04&   3 &   0.02$\pm$0.02&   3 &  -0.13$\pm$0.06&   3 &   0.11$\pm$0.04&   2 &  -0.35$\pm$0.02&   6 &  -0.72$\pm$0.04&   3 &  -0.19$\pm$0.04&   4 &   0.10$\pm$0.07&   2 &   0.11$\pm$0.08&   3 &   0.18$\pm$0.08\\
 ScI &   2 &  -0.01$\pm$0.10&   2 &   0.14$\pm$0.08&   1 &  -0.05         &   2 &   0.17$\pm$0.04& --  &    --          &   1 &  -1.73         &   1 &  -0.18         &   2 &   0.10$\pm$0.08&  -- &  --            &   2 &   0.15$\pm$0.07\\
 ScII&   3 &   0.02$\pm$0.09&   5 &   0.11$\pm$0.06&   6 &  -0.14$\pm$0.05&   5 &   0.11$\pm$0.06&   5 &  -0.34$\pm$0.07&   6 &  -0.96$\pm$0.05&   7 &  -0.20$\pm$0.04&   4 &   0.18$\pm$0.04&   3 &  -0.04$\pm$0.07&   7 &   0.15$\pm$0.06\\
 TiI &  21 &  -0.06$\pm$0.08&  32 &   0.04$\pm$0.08&  18 &  -0.23$\pm$0.07&  26 &   0.00$\pm$0.05&  16 &  -0.32$\pm$0.05&  22 &  -0.79$\pm$0.05&  22 &  -0.23$\pm$0.07&  21 &   0.05$\pm$0.06&  14 &   0.01$\pm$0.04&  25 &   0.06$\pm$0.06\\
 VI  &  18 &  -0.02$\pm$0.05&  24 &   0.09$\pm$0.09&   2 &  -0.24$\pm$0.01&  19 &   0.14$\pm$0.07&   3 &  -0.31$\pm$0.09&  16 &  -0.99$\pm$0.06&   7 &  -0.16$\pm$0.07&  18 &   0.10$\pm$0.06&   4 &   0.11$\pm$0.11&  19 &   0.14$\pm$0.07\\
 CrI &   6 &  -0.24$\pm$0.07&   6 &  -0.12$\pm$0.05&   8 &  -0.38$\pm$0.06&  12 &   0.12$\pm$0.07&   7 &  -0.70$\pm$0.08&   3 &  -1.17$\pm$0.03&   6 &  -0.47$\pm$0.05&   6 &  -0.21$\pm$0.07&   4 &  -0.20$\pm$0.09&  10 &   0.00$\pm$0.08\\
 MnI &   1 &  -0.25         &   1 &  -0.30         &   1 &  -0.69         &   1 &  -0.07         &   1 &  -0.91         &   2 &  -1.37$\pm$0.13&   1 &  -0.82         & --  &    --          &   1 &  -0.58         &   1 &  -0.10         \\
 FeI &  71 &  -0.15$\pm$0.07&  88 &  -0.09$\pm$0.07&  64 &  -0.30$\pm$0.07&  80 &   0.11$\pm$0.06&  34 &  -0.55$\pm$0.08&  55 &  -1.01$\pm$0.06&  61 &  -0.43$\pm$0.06&  73 &  -0.09$\pm$0.06&  51 &  -0.13$\pm$0.06&  77 &   0.05$\pm$0.06\\
 FeII&   5 &  -0.24$\pm$0.07&   6 &  -0.14$\pm$0.07&   4 &  -0.39$\pm$0.02&   7 &   0.09$\pm$0.05&   2 &  -0.68$\pm$0.06&   6 &  -1.08$\pm$0.03&   3 &  -0.46$\pm$0.02&   5 &  -0.16$\pm$0.05&   7 &  -0.25$\pm$0.08&   7 &   0.01$\pm$0.08\\
 CoI &   5 &  -0.19$\pm$0.09&   6 &  -0.07$\pm$0.10&   3 &  -0.33$\pm$0.02&   6 &  -0.02$\pm$0.14&   4 &  -0.55$\pm$0.04&   2 &  -1.02$\pm$0.07&   4 &  -0.39$\pm$0.04&   5 &  -0.05$\pm$0.07&   5 &  -0.10$\pm$0.11&   5 &  -0.03$\pm$0.07\\
 NiI &  21 &  -0.15$\pm$0.08&  26 &  -0.04$\pm$0.06&  20 &  -0.35$\pm$0.07&  22 &   0.20$\pm$0.06&   5 &  -0.60$\pm$0.04&  15 &  -1.06$\pm$0.03&  15 &  -0.42$\pm$0.06&  25 &  -0.06$\pm$0.06&  17 &  -0.19$\pm$0.06&  26 &   0.04$\pm$0.07\\
 YII & --  &    --          & --  &    --          &  -- &    --          &   1 &  -0.08         &   2 &  -0.77$\pm$0.11& --  &    --          & --  &    --          & --  &    --          & --  &    --          & --  &    --          \\
 ZrII& --  &    --          & --  &    --          &   1 &  -0.49         & --  &    --          &   1 &  -0.41         &   1 &  -0.52         & --  &    --          &   1 &   0.13         & --  &    --          & --  &    --          \\
 BaII& --  &    --          &   1 &   0.06         &   1 &  -0.22         &   1 &   0.03         &   1 &  -0.02         &   1 &  -0.41         &   1 &  -0.13         &   1 &   0.14         &   1 &   0.41         &   1 &   0.52         \\
 LaII&   1 &   0.34         & --  &    --          &   1 &  -0.32         & --  &    --          & --  &    --          &   1 &  -0.66         &   1 &  -0.40         & --  &    --          &   1 &   0.06         &   2 &   0.15$\pm$0.09\\
 NdII&   1 &   0.02         &   1 &   0.01         &   1 &  -0.10         &   1 &  -0.13         &   2 &  -0.32$\pm$0.12&   2 &  -0.78$\pm$0.09&   2 &  -0.14$\pm$0.15&   1 &   0.04         &   1 &   0.08         &   1 &   0.21         \\
 EuII& --  &    --          &   1 &   0.10         & --  &    --          &   1 &  -0.04         &   1 &  -0.15         &   1 &  -0.77         &   1 &  -0.12         &   1 &   0.15         &   1 &   0.08         &   1 &   0.34         \\
\hline
\end{tabular}
\end{sidewaystable*}

\begin{figure*}[t]
\setcaptionmargin{0mm}
\onelinecaptionsfalse 
\renewcommand\arraystretch{1.0}
\captionstyle{flushleft}
\hspace{-2cm}\includegraphics[width=12cm]{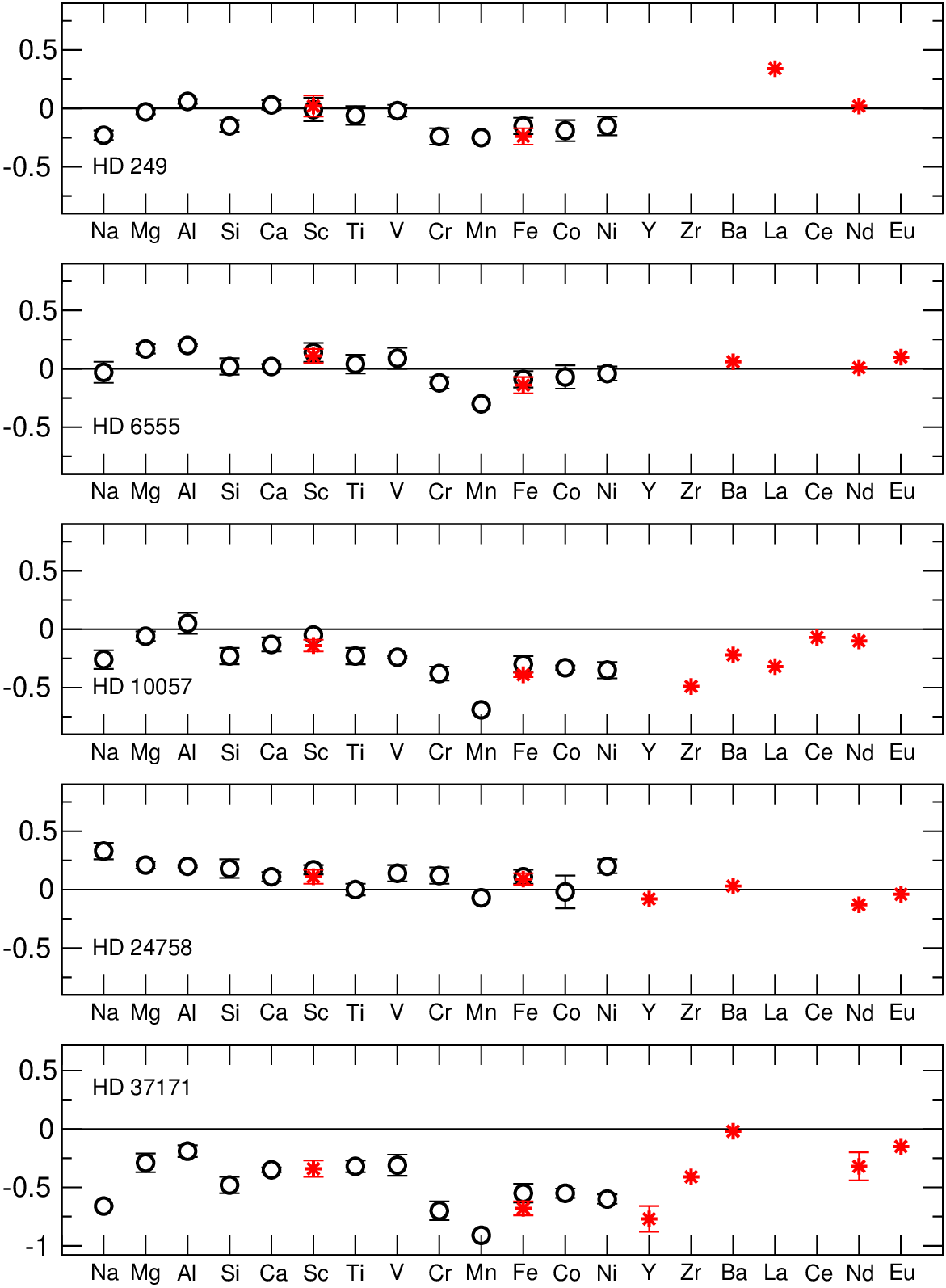}
\caption{\rm Relative abundances of chemical elements in atmospheres of the
studied stars. Abundances defined from neutral atoms marked by
circles, from ions marked by asterisks.}
\label{fig:abund}
\end{figure*}
 
\begin{figure*}[t]
\setcaptionmargin{0mm}
\onelinecaptionsfalse 
\renewcommand\arraystretch{1.0}
\captionstyle{flushleft}
\hspace{-2cm}\includegraphics[width=12cm]{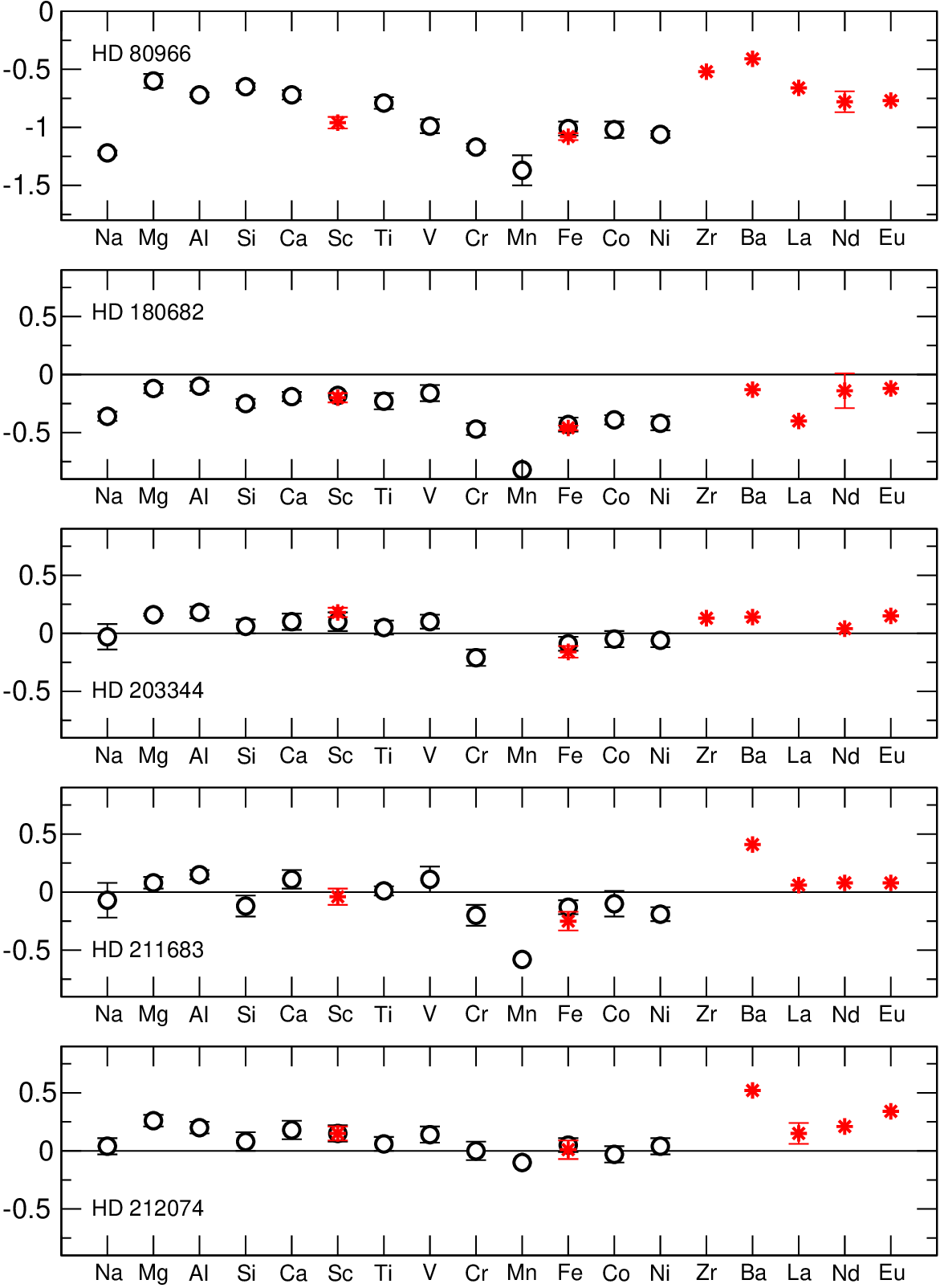}
\addtocounter{figure}{-1}
\caption{\rm continue}
\end{figure*}

\begin{table*}
\setcaptionmargin{0mm}
\renewcommand\arraystretch{0.7}
\onelinecaptionsfalse
\captionstyle{flushleft}
\caption{Changes in abundances in atmosphere of the star
HD37171 (\Teff=4000K \lgg=1.25 \Vt=1.35~\kms) and HD212074 (\Teff=4700K
\lgg=2.55 \Vt=1.30~\kms) due changes in $\Delta$\Teff=+100K,
$\Delta$\lgg=+0.10, $\Delta$\Vt=+0.10~\kms and total error $\Delta$}
\label{tab:abnerr}
\begin{tabular}{|c|r|r|r|r|r||r|r|r|r|r|}
\hline
Element&
N&\multicolumn{4}{c||}{$\Delta[El/H]_{HD37171}$}&
N&\multicolumn{4}{c|}{$\Delta[El/H]_{HD212074}$
} \\
& &
$\Delta$\Teff& 
$\Delta$\lgg& 
$\Delta$\Vt& 
$\Delta$ & &
$\Delta$\Teff& 
$\Delta$\lgg& 
$\Delta$\Vt& 
$\Delta$ \\
\hline
NA1 &   1&   0.10&  0.00& -0.02&  0.10&  2&   0.10&  0.00& -0.02&  0.10\\
MG1 &   2&  -0.02&  0.02& -0.01&  0.03&  2&   0.03&  0.00& -0.03&  0.04\\
AL1 &   2&   0.08&  0.00& -0.03&  0.09&  2&   0.08&  0.00& -0.02&  0.08\\
SI1 &   5&  -0.10&  0.04& -0.01&  0.11& 11&  -0.04&  0.03& -0.01&  0.05\\
CA1 &   2&   0.11&  0.00& -0.03&  0.11&  3&   0.08& -0.01& -0.02&  0.08\\
SC2 &   5&  -0.03&  0.04& -0.04&  0.06&  7&  -0.02&  0.04& -0.03&  0.05\\
TI1 &  16&   0.14&  0.01& -0.04&  0.15& 25&   0.14&  0.00& -0.03&  0.14\\
 V1 &   3&   0.14&  0.01& -0.03&  0.14& 19&   0.17&  0.01& -0.05&  0.18\\
CR1 &   7&   0.11&  0.01& -0.04&  0.12& 10&   0.10&  0.00& -0.03&  0.10\\
MN1 &   1&   0.03&  0.02& -0.01&  0.04&  1&   0.06&  0.01& -0.01&  0.06\\
FE1 &  34&  -0.02&  0.03& -0.04&  0.05& 77&   0.04&  0.01& -0.04&  0.06\\
FE2 &   2&  -0.21&  0.08& -0.03&  0.23&  7&  -0.11&  0.06& -0.03&  0.13\\
CO1 &   4&  -0.01&  0.03& -0.05&  0.06&  7&   0.06&  0.02& -0.05&  0.08\\
NI1 &   5&  -0.06&  0.03& -0.04&  0.08& 26&   0.01&  0.02& -0.04&  0.05\\
 Y2 &   2&  -0.03&  0.05& -0.01&  0.06&   &       &      &      &      \\
ZR2 &   1&  -0.03&  0.05&  0.00&  0.06&  2&   0.20&  0.01& -0.01&  0.20\\
BA2 &   1&   0.00&  0.00& -0.13&  0.13&  1&   0.02&  0.02& -0.10&  0.10\\
ND2 &   2&   0.04&  0.04& -0.03&  0.06&  1&   0.03&  0.04& -0.02&  0.05\\
EU2 &   1&   0.00&  0.05& -0.02&  0.05&  1&  -0.02&  0.03& -0.05&  0.06\\
\hline         
\end{tabular}
\end{table*}

\noindent
We determined the stellar atmosphere parameters using a technique based on
Kurucz's model atmospheres \cite{2003IAUS..210P.A20C} and analysis of the
relative abundances of iron-peak elements. The technique is described in detail
in \cite{2001ARep...45..301B}) and allows the stellar atmosphere parameters for
G-K giants to be determined with an accuracy of about 70-100~K for \Teff,
0.10-0.15 for \lgg, and 0.10-0.15~\kms for \Vt. For late-K giants, the
accuracy can be lower due to the greater influence of blending by atomic and
molecular lines. In this paper, when analyzing the relative abundances of
iron-peak elements when determining the stellar atmosphere parameters, we
disregarded titanium, because it is well known that the [Ti/Fe] abundance can
be enhanced at low metallicities and for thick-disk stars
\cite{2005A&A...433..185B}. Using the derived parameters (\Teff, \lgg, \Vt ),
we computed the corresponding model stellar atmospheres with the ATLAS9 code
\cite{1993KurCD..13.....K}.

Table~\ref{tab:param} gives our estimates of the stellar atmosphere parameters
(effective temperature \Teff , surface gravity \lgg, microturbulence \Vt , and
metallicity [Fe/H]), masses, and interstellar extinctions $A_V$ . We determined
the masses based on evolutionary tracks from \cite{2000A&AS..141..371G} by
taking into account the stellar metallicity. The interstellar extinctions were
estimated from the color excess E(B-V); the dereddened colors were calculated
from calibrations \cite{1998A&A...333..231B} based on Kurucz's model stellar
atmospheres. Based on the measured equivalent widths of the selected unblended
spectral lines, we estimated the elemental abundances with the WIDTH9 code.
These are presented in Table~\ref{tab:abund} and Fig.~\ref{fig:abund}, where the
open circles and asterisks denote the abundances determined from the spectral
lines of neutral and ionized atoms, respectively. The list of selected lines
with their characteristics and equivalent widths is available in electronic
form. The cobalt abundance was determined by taking into account the hyperfine
splitting effect, which can be strong in the case of cool giants
\cite{2008ARep...52..630B}. The abundance
errors given in Table~\ref{tab:abund} and marked by the bars in
Fig.~\ref{fig:abund} were determined as the dispersion of the individual
abundances calculated from individual spectral lines. As an example, the
possible abundance errors associated with the determination of stellar
atmosphere parameters are listed in Table~\ref{tab:abnerr} for two stars.
Table~\ref{tab:abnerr} gives the number of lines used ($N$) and the changes in
the abundance of each element when changing individual model parameters
($\Delta$\Teff=+100K, $\Delta$\lgg=+0.10, $\Delta$\Vt=+0.10~\kms) and
the total change in abundance $\Delta$.

\begin{figure*}
\setcaptionmargin{0mm}
\onelinecaptionsfalse 
\renewcommand\arraystretch{1.0}
\captionstyle{flushleft}
\hspace{-2cm}\includegraphics[width=12cm]{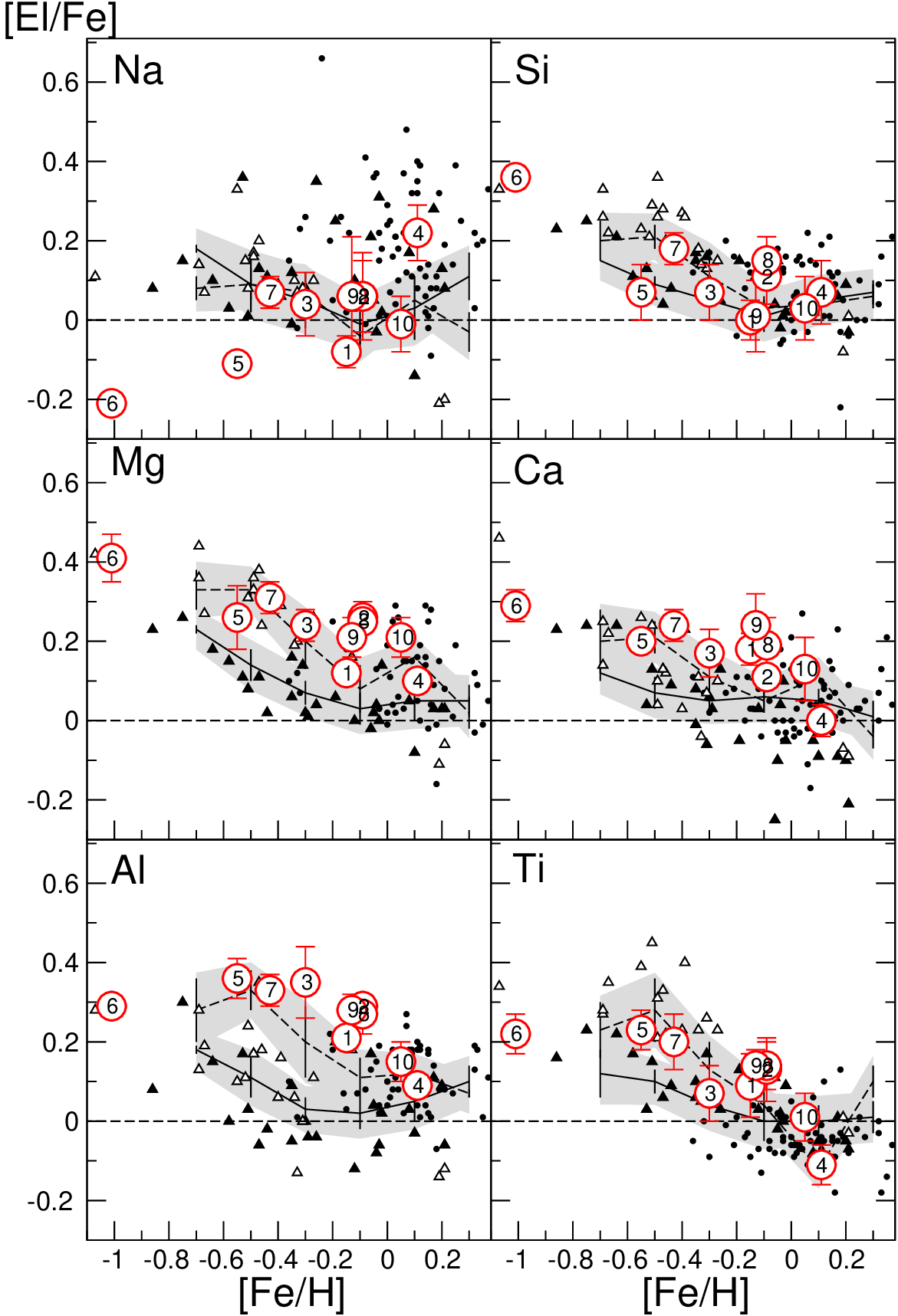}
\caption{\rm Trends of abundances of [Na,Mg,Al,Si,Ca,Ti/Fe] vs [Fe/H] in
atmospheres of the studied giants in comparison with data of other stars.
Early studied red giants of thin disk marked by small filled circles.
Filles and open triangles indicate on the reg giants from
\cite{2010A&A...513A..35A}. Continuous and dashed lines - the average
data of dwarfs of thin and thick disk published by \cite{2005A&A...433..185B},
with interval 0.2~dex in metallicity, filled area indicate$1\sigma$  dispersion
level.}
\label{fig:all}
\end{figure*}

The sodium abundance was determined from the 6154\AA\ and 6160~\AA lines without
any correction for non-LTE processes. According to \cite{1989SvA....33..449K,
2000ARep...44..790M, 2011A&A...528A.103L}, this doublet is formed deeper than
other sodium lines, and the non-LTE processes do not introduce significant
deviations in the abundance determination ($<$0.1~dex). Figure~\ref{fig:all}
presents the abundance trends for some elements with metallicity. The large
circles mark our thick-disk red giants with their ordinal numbers from
Table\ref{tab:list}. The small filled circles indicate the 74 thin-disk red
giants from \cite{2001ARep...45..700A, 2002ARep...46..819B,
2003ARep...47..648A, 2004ARep...48..597A, 2005ARep...49..535A,
2008ARep...52..630B, 2009ARep...53..660P, 2009ARep...53..685P,
2011ARep...55..256P}
that we studied previously by a unified technique. The filled and open
triangles indicate the thin-disk (29 stars) and thick-disk (22 stars) red giants
from \cite{2010A&A...513A..35A}, who investigated the Galactic chemical
evolution in the solar neighborhood. The solid and dashed lines indicate the
thin- and thick disk dwarfs from \cite{2005A&A...433..185B} averaged with a
metallicity interval of 0.2 dex, while the shaded region denote a dispersion of
$1\sigma$.

\section*{DISCUSSION}

\begin{figure*}
\setcaptionmargin{0mm}
\onelinecaptionsfalse 
\renewcommand\arraystretch{1.0}
\captionstyle{flushleft}
\hspace{-2cm}\includegraphics[width=12cm]{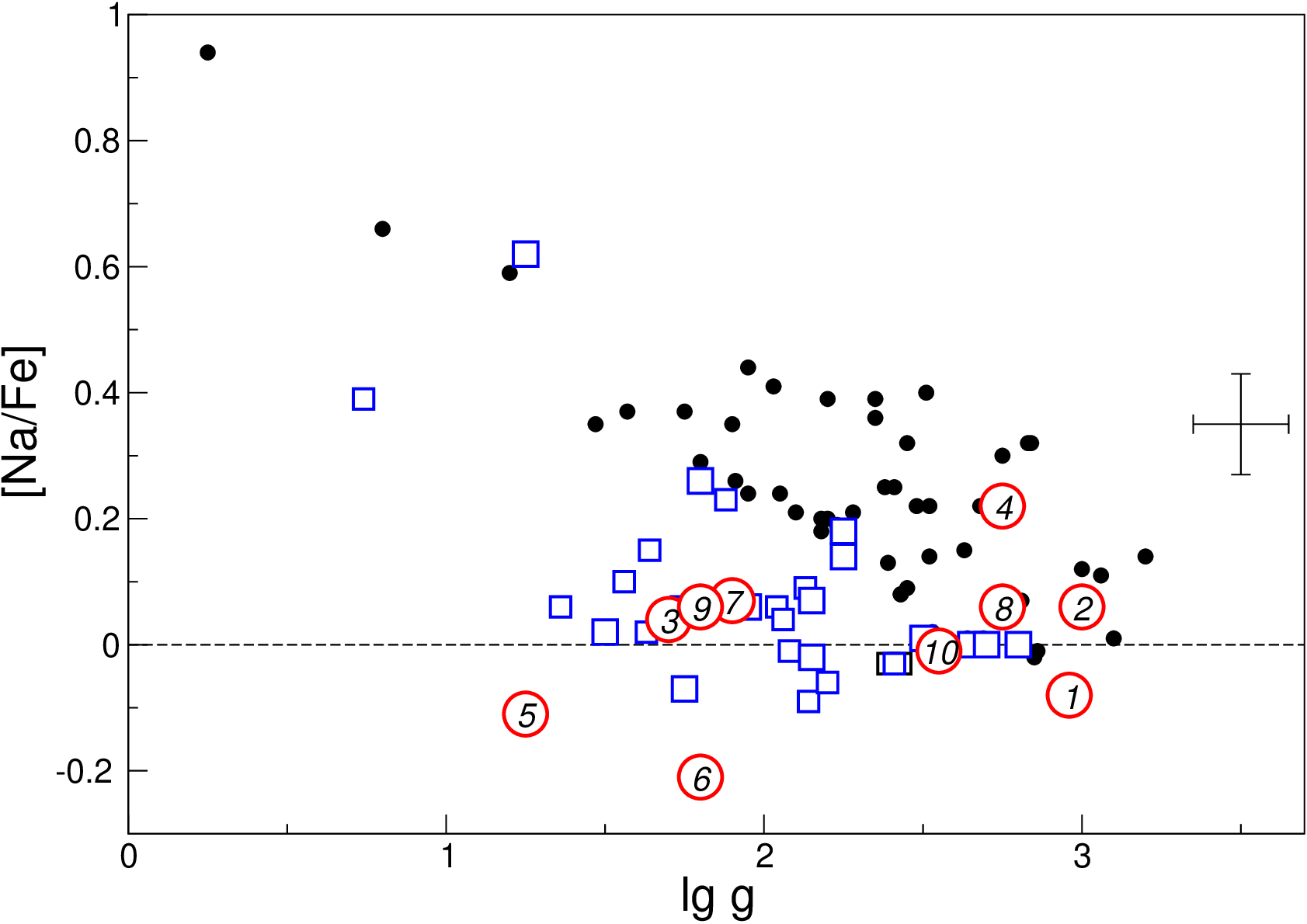}
\caption{\rm [Na/Fe] vs \lgg\ in atmospheres of the studied giants of
thin disk (filled circles) and thick disk (open rectangles) and red giants of
current study (big open circles). The typical error bar in right part.}
\label{fig:Na}
\end{figure*}

\noindent
Figure~\ref{fig:all} demonstrates a compact arrangement of stars for all
elements, except sodium, which reflects the evolution of elemental abundances
in the Galaxy. The separation of the trends constructed for the thin and thick
disks is characteristics of all elements, except sodium. Both dwarfs and giants
of each of the disks are located in the same regions, i.e., the abundances of
these elements do not change over the elapsed lifetime of the star. In the case
of sodium, a compact arrangement is observed only for dwarfs. The trends for
thin- and thick-disk stars coincide, within the error limits. In contrast, red
giants are chaotically scattered and no dependence on metallicity can be
distinguished. Such a behavior suggests that the [Na/Fe] abundance is determined
not only by the chemical evolution of the Galaxy but also by the evolution of
the star itself. The amount of sodium that was synthesized in the stellar core
at the main-sequence stage and that was brought by convective flows into the
stellar atmosphere is added to the initial abundance. The thick disk red giants
in Fig.~\ref{fig:all}, within the error limits, are located in the region of
dwarfs and exhibit no detectable sodium overabundances. However, this is most
likely the selection effect, because some of the thick-disk red giants from
\cite{2010A&A...513A..35A} and \cite{2012AstL...38..101P} have significant
[Na/Fe] overabundances. HD80966 with the lowest metallicity exhibits the
lowest (and atypical of the remaining stars) abundance [Na/Fe]=-0.21~dex with an
error of 0.02~dex calculated as the mean of the abundances from the two lines.
Analysis of the possible errors by taking into account the uncertainty in the
iron abundance and model parameters leads to a total error of about 0.12~dex. If
we explain the low abundance by the error in determining the temperature, then
it turns out that we underestimate the temperature approximately by 200~K. Such
a change will lead to a significant discrepancy between the elemental abundances
determined from lines with different lower level excitation potentials, while
the position of the star on other plots in Fig.~\ref{fig:all} will rise
dramatically. The discrepancy will also affect the [FeI/H] and [FeII/H]
abundances, which can be corrected by varying the surface gravity \lgg. However,
the value of \lgg determined here is in good agreement with \lgg\ derived from
the parallax (1.80$\pm$0.15 and 1.86$\pm$0.23, respectively). Nevertheless, the
thick-disk giants from our database, on average, exhibit a lower [Na/Fe]
abundance than the thin-disk ones: 0.03$\pm$0.10 and 0.17 б╠ 0.15, respectively.
\cite{2011AJ....141...58J}, who investigated the chemical composition of red
giants from two thick-disk open star clusters, NGC~2204 and NGC~2243, also
pointed out that at a metallicity [Fe/H] from -0.4 to -0.2 and \lgg\ from 1.5 to
2.9, the mean [Na/Fe] abundance is close to zero: 0.02$\pm$0.04. An increase in
the atmospheric [Na/Fe] abundance of red giants is better demonstrated by
Fig.~\ref{fig:Na}, which shows the dependence on surface gravity \lgg. Both
these quantities are determined directly from observations. In the figure, the
thin-disk stars (filled circles) show a rise as the surface gravity decreases
from \lgg$\approx$3-3.5. The surface gravity is related to other fundamental
stellar parameters by the relation \lgg = -10.607+lg\,$M$/\Mo+
4lg\Teff-lg$L$, where $M$/\Mo is the stellar mass in solar masses and lg$L$ is
the stellar luminosity. At a constant stellar mass, the decrease in \lgg\ can
be associated with a decrease in the effective temperature and an increase in
the stellar luminosity. Such a behavior is typical of a star as it evolves along
the red giant branch, and the appearance of sodium overabundances and the
increase in sodium abundance due to the rise of sodium-enriched matter through
convective flows \cite{2004MmSAI..75..347W} are associated precisely with this
evolutionary stage. At this stage of stellar evolution, an enrichment of matter
by sodium is possible only in the regions of nuclear reactions at the
main-sequence phase. The hydrogen burning reaction in the neon-sodium cycle is
the main source of sodium ~\cite{1996ApJ...464L..79C, 1998ApJ...492..575C}.
In the interiors of low-mass (1-3~$M_\odot$) stars, this cycle is not closed and
turns into the chain of reactions
$^{20}Ne(p,\gamma)^{21}Na(\beta^+)^{21}Ne(p,\gamma)^{22}Na(\beta^+)^{22}Ne(p,
\gamma)^{23}Na$
The first quantitative estimate of these reactions for supergiants is given in
\cite{1987SvAL...13..214D}: depending on the temperature, from 1/3 to 2/3 of
the $^{23}Na$ nuclei pass into $^{20}Ne$. At the same time, the sodium abundance
in the atmospheres of F-K supergiants increases by a factor of 5 (0.7 dex), in
agreement with the observational data and Fig.~\ref{fig:Na}. Various conditions
for the nuclear reactions of the neon-sodium cycle were also considered
by \cite{RevModPhys..69.995W} and \cite{1994PASJ...46..395T}.

The thick-disk red giants are marked in Fig.~\ref{fig:Na} by the large circles
(from this paper, the ordinal numbers correspond to Table~\ref{tab:list}) and
squares (previously investigated ones). It can be seen from the figure that they
are located systematically below the thin disk red giants, while for the region
\lgg$<$2 the separation between these two groups close in the number of stars
exceeds the [Na/Fe] abundance errors. Thus, the difference in [Na/Fe] abundance
in the atmospheres of thin- and thick-disk red giants is significant. Among the
stars whose parameters were entered into the database, there are several pairs
of thin- and thick-disk stars with similar parameters (\Teff , \lgg, [Fe/H])
(see \cite{2012AstL...38..101P}). Such stars also have similar masses and ages
but differ only by the [Na/Fe] abundance and membership in different kinematic
groups of stars in the Galaxy. Such a difference is not observed for the thin-
and thick-disk dwarfs (see Fig.~\ref{fig:all}). Consequently, the initial sodium
abundance must be also the same for those stars that have presently become red
giants, while the nature of the difference is in the production of sodium in the
NeNa cycle. A neon underabundance may exist in thick-disk objects. This is hard
to check, because the neon lines are difficult to observe. However, the main 20
Ne isotope is an $\alpha$-process element, and its behavior with metallicity
must follow the trends of other $\alpha$-elements, i.e., an increase in
abundance with decreasing metallicity \cite{2001A&A...370.1103A}. In this 
case, we should have seen higher overabundances of sodium formed from neon in
thick-disk giants, but the reverse is true. The $^{22}Ne$ underabundance is a
possible cause. Indeed, \cite{RevModPhys..69.995W} point out a signifficant
reprocessing of $^{22}Ne$ into $^{23}Na$, while the initial $^{22}Ne$/$^{23}Na$
abundance ratio in dwarfs is about 4.1. In any case, the difference between the
atmospheric [Na/Fe] abundances of thin- and thick disk red giants requires an
explanation and invoking theoretical calculations.

\section*{CONCLUSIONS}
\noindent
Thus, based on spectroscopic observations, we determined the stellar atmosphere
parameters for ten thick-disk red giants. We confirmed the difference in
atmospheric [Na/Fe] abundance between thin- and thick-disk giants. Its nature
may be related to the neon isotope abundance difference and this hypothesis
requires verification.

\textit{Acknowledgments}
\noindent
This work was supported by the Program for Support of Leading Scientific
Schools (project 3602.2012.2), the ``Nonstationary Phenomena in Objects of the
Universe'' Program of the Presidium of the Russian Academy of Sciences, and the
Federal Agency for Science and Innovations (2012-1.5-12000-1011-014
``Influence of the Formation of Exoplanets on the Observed Properties of Outer
Stellar Layers'')

\vskip 5mm 
\bibliographystyle{aipauth4-1}
\bibliography{paper}

\begin{thebibliography}{49}%
\makeatletter
\providecommand \@ifxundefined [1]{%
 \@ifx{#1\undefined}
}%
\providecommand \@ifnum [1]{%
 \ifnum #1\expandafter \@firstoftwo
 \else \expandafter \@secondoftwo
 \fi
}%
\providecommand \@ifx [1]{%
 \ifx #1\expandafter \@firstoftwo
 \else \expandafter \@secondoftwo
 \fi
}%
\providecommand \natexlab [1]{#1}%
\providecommand \enquote  [1]{``#1''}%
\providecommand \bibnamefont  [1]{#1}%
\providecommand \bibfnamefont [1]{#1}%
\providecommand \citenamefont [1]{#1}%
\providecommand \href@noop [0]{\@secondoftwo}%
\providecommand \href [0]{\begingroup \@sanitize@url \@href}%
\providecommand \@href[1]{\@@startlink{#1}\@@href}%
\providecommand \@@href[1]{\endgroup#1\@@endlink}%
\providecommand \@sanitize@url [0]{\catcode `\\12\catcode `\$12\catcode
  `\&12\catcode `\#12\catcode `\^12\catcode `\_12\catcode `\%12\relax}%
\providecommand \@@startlink[1]{}%
\providecommand \@@endlink[0]{}%
\providecommand \url  [0]{\begingroup\@sanitize@url \@url }%
\providecommand \@url [1]{\endgroup\@href {#1}{\urlprefix }}%
\providecommand \urlprefix  [0]{URL }%
\providecommand \Eprint [0]{\href }%
\providecommand \doibase [0]{http://dx.doi.org/}%
\providecommand \selectlanguage [0]{\@gobble}%
\providecommand \bibinfo  [0]{\@secondoftwo}%
\providecommand \bibfield  [0]{\@secondoftwo}%
\providecommand \translation [1]{[#1]}%
\providecommand \BibitemOpen [0]{}%
\providecommand \bibitemStop [0]{}%
\providecommand \bibitemNoStop [0]{.\EOS\space}%
\providecommand \EOS [0]{\spacefactor3000\relax}%
\providecommand \BibitemShut  [1]{\csname bibitem#1\endcsname}%
\let\auto@bib@innerbib\@empty
\bibitem [{\citenamefont {{Alib{\'e}s}}, \citenamefont {{Labay}},\ and\
  \citenamefont {{Canal}}(2001)}]{2001A&A...370.1103A}%
  \BibitemOpen
  \bibfield  {author} {\bibinfo {author} {\bibnamefont {{Alib{\'e}s}},
  \bibfnamefont {A.}}, \bibinfo {author} {\bibnamefont {{Labay}}, \bibfnamefont
  {J.}}, \ and\ \bibinfo {author} {\bibnamefont {{Canal}}, \bibfnamefont
  {R.}},\ }\href {\doibase 10.1051/0004-6361:20010296} {\bibfield  {journal}
  {\bibinfo  {journal} {Astron. and Astrophys.}\ }\textbf {\bibinfo {volume}
  {370}},\ \bibinfo {pages} {1103} (\bibinfo {year} {2001})},\ \Eprint
  {http://arxiv.org/abs/arXiv:astro-ph/0012505} {arXiv:astro-ph/0012505}
  \BibitemShut {NoStop}%
\bibitem [{\citenamefont {{Allen}}\ and\ \citenamefont
  {{Santillan}}(1991)}]{1991RMxAA..22..255A}%
  \BibitemOpen
  \bibfield  {author} {\bibinfo {author} {\bibnamefont {{Allen}}, \bibfnamefont
  {C.}}\ and\ \bibinfo {author} {\bibnamefont {{Santillan}}, \bibfnamefont
  {A.}},\ }\href@noop {} {\bibfield  {journal} {\bibinfo  {journal} {Revista
  Mexicana de Astronomia y Astrofisica}\ }\textbf {\bibinfo {volume} {22}},\
  \bibinfo {pages} {255} (\bibinfo {year} {1991})}\BibitemShut {NoStop}%
\bibitem [{\citenamefont {{Alves-Brito}}\ \emph {et~al.}(2010)\citenamefont
  {{Alves-Brito}}, \citenamefont {{Mel{\'e}ndez}}, \citenamefont {{Asplund}},
  \citenamefont {{Ram{\'{\i}}rez}},\ and\ \citenamefont
  {{Yong}}}]{2010A&A...513A..35A}%
  \BibitemOpen
  \bibfield  {author} {\bibinfo {author} {\bibnamefont {{Alves-Brito}},
  \bibfnamefont {A.}}, \bibinfo {author} {\bibnamefont {{Mel{\'e}ndez}},
  \bibfnamefont {J.}}, \bibinfo {author} {\bibnamefont {{Asplund}},
  \bibfnamefont {M.}}, \bibinfo {author} {\bibnamefont {{Ram{\'{\i}}rez}},
  \bibfnamefont {I.}}, \ and\ \bibinfo {author} {\bibnamefont {{Yong}},
  \bibfnamefont {D.}},\ }\href {\doibase 10.1051/0004-6361/200913444}
  {\bibfield  {journal} {\bibinfo  {journal} {Astron. and Astrophys.}\ }\textbf
  {\bibinfo {volume} {513}},\ \bibinfo {pages} {A35+} (\bibinfo {year}
  {2010})},\ \Eprint {http://arxiv.org/abs/1001.2521} {arXiv:1001.2521
  [astro-ph.SR]} \BibitemShut {NoStop}%
\bibitem [{\citenamefont {{Antipova}}\ and\ \citenamefont
  {{Boyarchuk}}(2001)}]{2001ARep...45..700A}%
  \BibitemOpen
  \bibfield  {author} {\bibinfo {author} {\bibnamefont {{Antipova}},
  \bibfnamefont {L.~I.}}\ and\ \bibinfo {author} {\bibnamefont {{Boyarchuk}},
  \bibfnamefont {A.~A.}},\ }\href {\doibase 10.1134/1.1398919} {\bibfield
  {journal} {\bibinfo  {journal} {Astronomy Reports}\ }\textbf {\bibinfo
  {volume} {45}},\ \bibinfo {pages} {700} (\bibinfo {year} {2001})}\BibitemShut
  {NoStop}%
\bibitem [{\citenamefont {{Antipova}}\ \emph {et~al.}(2003)\citenamefont
  {{Antipova}}, \citenamefont {{Boyarchuk}}, \citenamefont {{Pakhomov}},\ and\
  \citenamefont {{Panchuk}}}]{2003ARep...47..648A}%
  \BibitemOpen
  \bibfield  {author} {\bibinfo {author} {\bibnamefont {{Antipova}},
  \bibfnamefont {L.~I.}}, \bibinfo {author} {\bibnamefont {{Boyarchuk}},
  \bibfnamefont {A.~A.}}, \bibinfo {author} {\bibnamefont {{Pakhomov}},
  \bibfnamefont {Y.~V.}}, \ and\ \bibinfo {author} {\bibnamefont {{Panchuk}},
  \bibfnamefont {V.~E.}},\ }\href {\doibase 10.1134/1.1601633} {\bibfield
  {journal} {\bibinfo  {journal} {Astronomy Reports}\ }\textbf {\bibinfo
  {volume} {47}},\ \bibinfo {pages} {648} (\bibinfo {year} {2003})}\BibitemShut
  {NoStop}%
\bibitem [{\citenamefont {{Antipova}}\ \emph {et~al.}(2004)\citenamefont
  {{Antipova}}, \citenamefont {{Boyarchuk}}, \citenamefont {{Pakhomov}},\ and\
  \citenamefont {{Panchuk}}}]{2004ARep...48..597A}%
  \BibitemOpen
  \bibfield  {author} {\bibinfo {author} {\bibnamefont {{Antipova}},
  \bibfnamefont {L.~I.}}, \bibinfo {author} {\bibnamefont {{Boyarchuk}},
  \bibfnamefont {A.~A.}}, \bibinfo {author} {\bibnamefont {{Pakhomov}},
  \bibfnamefont {Y.~V.}}, \ and\ \bibinfo {author} {\bibnamefont {{Panchuk}},
  \bibfnamefont {V.~E.}},\ }\href {\doibase 10.1134/1.1777277} {\bibfield
  {journal} {\bibinfo  {journal} {Astronomy Reports}\ }\textbf {\bibinfo
  {volume} {48}},\ \bibinfo {pages} {597} (\bibinfo {year} {2004})}\BibitemShut
  {NoStop}%
\bibitem [{\citenamefont {{Antipova}}\ \emph {et~al.}(2005)\citenamefont
  {{Antipova}}, \citenamefont {{Boyarchuk}}, \citenamefont {{Pakhomov}},\ and\
  \citenamefont {{Yushkin}}}]{2005ARep...49..535A}%
  \BibitemOpen
  \bibfield  {author} {\bibinfo {author} {\bibnamefont {{Antipova}},
  \bibfnamefont {L.~I.}}, \bibinfo {author} {\bibnamefont {{Boyarchuk}},
  \bibfnamefont {A.~A.}}, \bibinfo {author} {\bibnamefont {{Pakhomov}},
  \bibfnamefont {Y.~V.}}, \ and\ \bibinfo {author} {\bibnamefont {{Yushkin}},
  \bibfnamefont {M.~V.}},\ }\href {\doibase 10.1134/1.1985951} {\bibfield
  {journal} {\bibinfo  {journal} {Astronomy Reports}\ }\textbf {\bibinfo
  {volume} {49}},\ \bibinfo {pages} {535} (\bibinfo {year} {2005})}\BibitemShut
  {NoStop}%
\bibitem [{\citenamefont {{Bensby}}\ \emph {et~al.}(2005)\citenamefont
  {{Bensby}}, \citenamefont {{Feltzing}}, \citenamefont {{Lundstr{\"o}m}},\
  and\ \citenamefont {{Ilyin}}}]{2005A&A...433..185B}%
  \BibitemOpen
  \bibfield  {author} {\bibinfo {author} {\bibnamefont {{Bensby}},
  \bibfnamefont {T.}}, \bibinfo {author} {\bibnamefont {{Feltzing}},
  \bibfnamefont {S.}}, \bibinfo {author} {\bibnamefont {{Lundstr{\"o}m}},
  \bibfnamefont {I.}}, \ and\ \bibinfo {author} {\bibnamefont {{Ilyin}},
  \bibfnamefont {I.}},\ }\href {\doibase 10.1051/0004-6361:20040332} {\bibfield
   {journal} {\bibinfo  {journal} {Astron. \& Astrophys.}\ }\textbf {\bibinfo
  {volume} {433}},\ \bibinfo {pages} {185} (\bibinfo {year}
  {2005})}\BibitemShut {NoStop}%
\bibitem [{\citenamefont {{Bessell}}, \citenamefont {{Castelli}},\ and\
  \citenamefont {{Plez}}(1998)}]{1998A&A...333..231B}%
  \BibitemOpen
  \bibfield  {author} {\bibinfo {author} {\bibnamefont {{Bessell}},
  \bibfnamefont {M.~S.}}, \bibinfo {author} {\bibnamefont {{Castelli}},
  \bibfnamefont {F.}}, \ and\ \bibinfo {author} {\bibnamefont {{Plez}},
  \bibfnamefont {B.}},\ }\href@noop {} {\bibfield  {journal} {\bibinfo
  {journal} {Astron. \& Astrophys.}\ }\textbf {\bibinfo {volume} {333}},\
  \bibinfo {pages} {231} (\bibinfo {year} {1998})}\BibitemShut {NoStop}%
\bibitem [{\citenamefont {{Boyarchuk}}\ \emph {et~al.}(2001)\citenamefont
  {{Boyarchuk}}, \citenamefont {{Antipova}}, \citenamefont {{Boyarchuk}},\ and\
  \citenamefont {{Savanov}}}]{2001ARep...45..301B}%
  \BibitemOpen
  \bibfield  {author} {\bibinfo {author} {\bibnamefont {{Boyarchuk}},
  \bibfnamefont {A.~A.}}, \bibinfo {author} {\bibnamefont {{Antipova}},
  \bibfnamefont {L.~I.}}, \bibinfo {author} {\bibnamefont {{Boyarchuk}},
  \bibfnamefont {M.~E.}}, \ and\ \bibinfo {author} {\bibnamefont {{Savanov}},
  \bibfnamefont {I.~S.}},\ }\href {\doibase 10.1134/1.1361322} {\bibfield
  {journal} {\bibinfo  {journal} {Astronomy Reports}\ }\textbf {\bibinfo
  {volume} {45}},\ \bibinfo {pages} {301} (\bibinfo {year} {2001})}\BibitemShut
  {NoStop}%
\bibitem [{\citenamefont {{Boyarchuk}}, \citenamefont {{Antipova}},\ and\
  \citenamefont {{Pakhomov}}(2008)}]{2008ARep...52..630B}%
  \BibitemOpen
  \bibfield  {author} {\bibinfo {author} {\bibnamefont {{Boyarchuk}},
  \bibfnamefont {A.~A.}}, \bibinfo {author} {\bibnamefont {{Antipova}},
  \bibfnamefont {L.~I.}}, \ and\ \bibinfo {author} {\bibnamefont {{Pakhomov}},
  \bibfnamefont {Y.~V.}},\ }\href {\doibase 10.1134/S1063772908080040}
  {\bibfield  {journal} {\bibinfo  {journal} {Astronomy Reports}\ }\textbf
  {\bibinfo {volume} {52}},\ \bibinfo {pages} {630} (\bibinfo {year}
  {2008})}\BibitemShut {NoStop}%
\bibitem [{\citenamefont {{Boyarchuk}}\ and\ \citenamefont
  {{Boyarchuk}}(1981)}]{1981BCrAO..63...68B}%
  \BibitemOpen
  \bibfield  {author} {\bibinfo {author} {\bibnamefont {{Boyarchuk}},
  \bibfnamefont {A.~A.}}\ and\ \bibinfo {author} {\bibnamefont {{Boyarchuk}},
  \bibfnamefont {M.~E.}},\ }\href@noop {} {\bibfield  {journal} {\bibinfo
  {journal} {Bulletin Crimean Astrophysical Observatory}\ }\textbf {\bibinfo
  {volume} {63}},\ \bibinfo {pages} {68} (\bibinfo {year} {1981})}\BibitemShut
  {NoStop}%
\bibitem [{\citenamefont {{Boyarchuk}}\ and\ \citenamefont
  {{Lyubimkov}}(1981)}]{1981BCrAO..64....1B}%
  \BibitemOpen
  \bibfield  {author} {\bibinfo {author} {\bibnamefont {{Boyarchuk}},
  \bibfnamefont {A.~A.}}\ and\ \bibinfo {author} {\bibnamefont {{Lyubimkov}},
  \bibfnamefont {L.~S.}},\ }\href@noop {} {\bibfield  {journal} {\bibinfo
  {journal} {Bulletin Crimean Astrophysical Observatory}\ }\textbf {\bibinfo
  {volume} {64}},\ \bibinfo {pages} {1} (\bibinfo {year} {1981})}\BibitemShut
  {NoStop}%
\bibitem [{\citenamefont {{Boyarchuk}}\ and\ \citenamefont
  {{Lyubimkov}}(1983)}]{1983BCrAO..66..119B}%
  \BibitemOpen
  \bibfield  {author} {\bibinfo {author} {\bibnamefont {{Boyarchuk}},
  \bibfnamefont {A.~A.}}\ and\ \bibinfo {author} {\bibnamefont {{Lyubimkov}},
  \bibfnamefont {L.~S.}},\ }\href@noop {} {\bibfield  {journal} {\bibinfo
  {journal} {Bulletin Crimean Astrophysical Observatory}\ }\textbf {\bibinfo
  {volume} {66}},\ \bibinfo {pages} {119} (\bibinfo {year} {1983})}\BibitemShut
  {NoStop}%
\bibitem [{\citenamefont {{Boyarchuk}}\ \emph {et~al.}(2002)\citenamefont
  {{Boyarchuk}}, \citenamefont {{Pakhomov}}, \citenamefont {{Antipova}},\ and\
  \citenamefont {{Boyarchuk}}}]{2002ARep...46..819B}%
  \BibitemOpen
  \bibfield  {author} {\bibinfo {author} {\bibnamefont {{Boyarchuk}},
  \bibfnamefont {A.~A.}}, \bibinfo {author} {\bibnamefont {{Pakhomov}},
  \bibfnamefont {Y.~V.}}, \bibinfo {author} {\bibnamefont {{Antipova}},
  \bibfnamefont {L.~I.}}, \ and\ \bibinfo {author} {\bibnamefont {{Boyarchuk}},
  \bibfnamefont {M.~E.}},\ }\href {\doibase 10.1134/1.1515093} {\bibfield
  {journal} {\bibinfo  {journal} {Astronomy Reports}\ }\textbf {\bibinfo
  {volume} {46}},\ \bibinfo {pages} {819} (\bibinfo {year} {2002})}\BibitemShut
  {NoStop}%
\bibitem [{\citenamefont {{Castelli}}\ and\ \citenamefont
  {{Kurucz}}(2003)}]{2003IAUS..210P.A20C}%
  \BibitemOpen
  \bibfield  {author} {\bibinfo {author} {\bibnamefont {{Castelli}},
  \bibfnamefont {F.}}\ and\ \bibinfo {author} {\bibnamefont {{Kurucz}},
  \bibfnamefont {R.~L.}},\ }in\ \href@noop {} {\emph {\bibinfo {booktitle}
  {Modelling of Stellar Atmospheres}}},\ \bibinfo {series} {IAU Symposium},
  Vol.\ \bibinfo {volume} {210},\ \bibinfo {editor} {edited by\ \bibinfo
  {editor} {\bibfnamefont {N.}~\bibnamefont {{Piskunov}}}, \bibinfo {editor}
  {\bibfnamefont {W.~W.}\ \bibnamefont {{Weiss}}}, \ and\ \bibinfo {editor}
  {\bibfnamefont {D.~F.}\ \bibnamefont {{Gray}}}}\ (\bibinfo {year} {2003})\
  pp.\ \bibinfo {pages} {20P--+}\BibitemShut {NoStop}%
\bibitem [{\citenamefont {{Cavallo}}, \citenamefont {{Sweigart}},\ and\
  \citenamefont {{Bell}}(1996)}]{1996ApJ...464L..79C}%
  \BibitemOpen
  \bibfield  {author} {\bibinfo {author} {\bibnamefont {{Cavallo}},
  \bibfnamefont {R.~M.}}, \bibinfo {author} {\bibnamefont {{Sweigart}},
  \bibfnamefont {A.~V.}}, \ and\ \bibinfo {author} {\bibnamefont {{Bell}},
  \bibfnamefont {R.~A.}},\ }\href@noop {} {\bibfield  {journal} {\bibinfo
  {journal} {Astrophys. J.}\ }\textbf {\bibinfo {volume} {464}},\ \bibinfo
  {pages} {L79+} (\bibinfo {year} {1996})}\BibitemShut {NoStop}%
\bibitem [{\citenamefont {{Cavallo}}, \citenamefont {{Sweigart}},\ and\
  \citenamefont {{Bell}}(1998)}]{1998ApJ...492..575C}%
  \BibitemOpen
  \bibfield  {author} {\bibinfo {author} {\bibnamefont {{Cavallo}},
  \bibfnamefont {R.~M.}}, \bibinfo {author} {\bibnamefont {{Sweigart}},
  \bibfnamefont {A.~V.}}, \ and\ \bibinfo {author} {\bibnamefont {{Bell}},
  \bibfnamefont {R.~A.}},\ }\href@noop {} {\bibfield  {journal} {\bibinfo
  {journal} {Astrophys. J.}\ }\textbf {\bibinfo {volume} {492}},\ \bibinfo
  {pages} {575} (\bibinfo {year} {1998})}\BibitemShut {NoStop}%
\bibitem [{\citenamefont {{Cayrel}}\ and\ \citenamefont
  {{Cayrel}}(1963)}]{1963ApJ...137..431C}%
  \BibitemOpen
  \bibfield  {author} {\bibinfo {author} {\bibnamefont {{Cayrel}},
  \bibfnamefont {G.}}\ and\ \bibinfo {author} {\bibnamefont {{Cayrel}},
  \bibfnamefont {R.}},\ }\href {\doibase 10.1086/147521} {\bibfield  {journal}
  {\bibinfo  {journal} {Astrophys. J.}\ }\textbf {\bibinfo {volume} {137}},\
  \bibinfo {pages} {431} (\bibinfo {year} {1963})}\BibitemShut {NoStop}%
\bibitem [{\citenamefont {{Cayrel de Strobel}}\ \emph
  {et~al.}(1970)\citenamefont {{Cayrel de Strobel}}, \citenamefont
  {{Chauve-Godard}}, \citenamefont {{Hernandez}},\ and\ \citenamefont
  {{Vaziaga}}}]{1970A&A.....7..408C}%
  \BibitemOpen
  \bibfield  {author} {\bibinfo {author} {\bibnamefont {{Cayrel de Strobel}},
  \bibfnamefont {G.}}, \bibinfo {author} {\bibnamefont {{Chauve-Godard}},
  \bibfnamefont {J.}}, \bibinfo {author} {\bibnamefont {{Hernandez}},
  \bibfnamefont {G.}}, \ and\ \bibinfo {author} {\bibnamefont {{Vaziaga}},
  \bibfnamefont {M.~J.}},\ }\href@noop {} {\bibfield  {journal} {\bibinfo
  {journal} {Astron. \& Astrophys.}\ }\textbf {\bibinfo {volume} {7}},\
  \bibinfo {pages} {408} (\bibinfo {year} {1970})}\BibitemShut {NoStop}%
\bibitem [{\citenamefont {{Chen}}\ \emph {et~al.}(2001)\citenamefont {{Chen}},
  \citenamefont {{Stoughton}}, \citenamefont {{Smith}}, \citenamefont
  {{Uomoto}}, \citenamefont {{Pier}}, \citenamefont {{Yanny}}, \citenamefont
  {{Ivezi{\'c}}}, \citenamefont {{York}}, \citenamefont {{Anderson}},
  \citenamefont {{Annis}}, \citenamefont {{Brinkmann}}, \citenamefont
  {{Csabai}}, \citenamefont {{Fukugita}}, \citenamefont {{Hindsley}},
  \citenamefont {{Lupton}}, \citenamefont {{Munn}},\ and\ \citenamefont {{the
  SDSS Collaboration}}}]{2001ApJ...553..184C}%
  \BibitemOpen
  \bibfield  {author} {\bibinfo {author} {\bibnamefont {{Chen}}, \bibfnamefont
  {B.}}, \bibinfo {author} {\bibnamefont {{Stoughton}}, \bibfnamefont {C.}},
  \bibinfo {author} {\bibnamefont {{Smith}}, \bibfnamefont {J.~A.}}, \bibinfo
  {author} {\bibnamefont {{Uomoto}}, \bibfnamefont {A.}}, \bibinfo {author}
  {\bibnamefont {{Pier}}, \bibfnamefont {J.~R.}}, \bibinfo {author}
  {\bibnamefont {{Yanny}}, \bibfnamefont {B.}}, \bibinfo {author} {\bibnamefont
  {{Ivezi{\'c}}}, \bibfnamefont {{\v Z}.}}, \bibinfo {author} {\bibnamefont
  {{York}}, \bibfnamefont {D.~G.}}, \bibinfo {author} {\bibnamefont
  {{Anderson}}, \bibfnamefont {J.~E.}}, \bibinfo {author} {\bibnamefont
  {{Annis}}, \bibfnamefont {J.}}, \bibinfo {author} {\bibnamefont
  {{Brinkmann}}, \bibfnamefont {J.}}, \bibinfo {author} {\bibnamefont
  {{Csabai}}, \bibfnamefont {I.}}, \bibinfo {author} {\bibnamefont
  {{Fukugita}}, \bibfnamefont {M.}}, \bibinfo {author} {\bibnamefont
  {{Hindsley}}, \bibfnamefont {R.}}, \bibinfo {author} {\bibnamefont
  {{Lupton}}, \bibfnamefont {R.}}, \bibinfo {author} {\bibnamefont {{Munn}},
  \bibfnamefont {J.~A.}}, \ and\ \bibinfo {author} {\bibnamefont {{the SDSS
  Collaboration}},},\ }\href {\doibase 10.1086/320647} {\bibfield  {journal}
  {\bibinfo  {journal} {Astrophys. J.}\ }\textbf {\bibinfo {volume} {553}},\
  \bibinfo {pages} {184} (\bibinfo {year} {2001})}\BibitemShut {NoStop}%
\bibitem [{\citenamefont {{Denisenkov}}(1988)}]{1988SvAL...14..435D}%
  \BibitemOpen
  \bibfield  {author} {\bibinfo {author} {\bibnamefont {{Denisenkov}},
  \bibfnamefont {P.~A.}},\ }\href@noop {} {\bibfield  {journal} {\bibinfo
  {journal} {Soviet Astronomy Letters}\ }\textbf {\bibinfo {volume} {14}},\
  \bibinfo {pages} {435} (\bibinfo {year} {1988})}\BibitemShut {NoStop}%
\bibitem [{\citenamefont {{Denisenkov}}\ and\ \citenamefont
  {{Ivanov}}(1987)}]{1987SvAL...13..214D}%
  \BibitemOpen
  \bibfield  {author} {\bibinfo {author} {\bibnamefont {{Denisenkov}},
  \bibfnamefont {P.~A.}}\ and\ \bibinfo {author} {\bibnamefont {{Ivanov}},
  \bibfnamefont {V.~V.}},\ }\href@noop {} {\bibfield  {journal} {\bibinfo
  {journal} {Soviet Astronomy Letters}\ }\textbf {\bibinfo {volume} {13}},\
  \bibinfo {pages} {214} (\bibinfo {year} {1987})}\BibitemShut {NoStop}%
\bibitem [{\citenamefont {{Famaey}}\ \emph {et~al.}(2005)\citenamefont
  {{Famaey}}, \citenamefont {{Jorissen}}, \citenamefont {{Luri}}, \citenamefont
  {{Mayor}}, \citenamefont {{Udry}}, \citenamefont {{Dejonghe}},\ and\
  \citenamefont {{Turon}}}]{2005A&A...430..165F}%
  \BibitemOpen
  \bibfield  {author} {\bibinfo {author} {\bibnamefont {{Famaey}},
  \bibfnamefont {B.}}, \bibinfo {author} {\bibnamefont {{Jorissen}},
  \bibfnamefont {A.}}, \bibinfo {author} {\bibnamefont {{Luri}}, \bibfnamefont
  {X.}}, \bibinfo {author} {\bibnamefont {{Mayor}}, \bibfnamefont {M.}},
  \bibinfo {author} {\bibnamefont {{Udry}}, \bibfnamefont {S.}}, \bibinfo
  {author} {\bibnamefont {{Dejonghe}}, \bibfnamefont {H.}}, \ and\ \bibinfo
  {author} {\bibnamefont {{Turon}}, \bibfnamefont {C.}},\ }\href {\doibase
  10.1051/0004-6361:20041272} {\bibfield  {journal} {\bibinfo  {journal}
  {Astron. and Astrophys.}\ }\textbf {\bibinfo {volume} {430}},\ \bibinfo
  {pages} {165} (\bibinfo {year} {2005})}\BibitemShut {NoStop}%
\bibitem [{\citenamefont {{Gilmore}}\ and\ \citenamefont
  {{Reid}}(1983)}]{1983MNRAS.202.1025G}%
  \BibitemOpen
  \bibfield  {author} {\bibinfo {author} {\bibnamefont {{Gilmore}},
  \bibfnamefont {G.}}\ and\ \bibinfo {author} {\bibnamefont {{Reid}},
  \bibfnamefont {N.}},\ }\href@noop {} {\bibfield  {journal} {\bibinfo
  {journal} {Monthly Notices Roy. Astron. Soc.}\ }\textbf {\bibinfo {volume}
  {202}},\ \bibinfo {pages} {1025} (\bibinfo {year} {1983})}\BibitemShut
  {NoStop}%
\bibitem [{\citenamefont {{Girardi}}\ \emph {et~al.}(2000)\citenamefont
  {{Girardi}}, \citenamefont {{Bressan}}, \citenamefont {{Bertelli}},\ and\
  \citenamefont {{Chiosi}}}]{2000A&AS..141..371G}%
  \BibitemOpen
  \bibfield  {author} {\bibinfo {author} {\bibnamefont {{Girardi}},
  \bibfnamefont {L.}}, \bibinfo {author} {\bibnamefont {{Bressan}},
  \bibfnamefont {A.}}, \bibinfo {author} {\bibnamefont {{Bertelli}},
  \bibfnamefont {G.}}, \ and\ \bibinfo {author} {\bibnamefont {{Chiosi}},
  \bibfnamefont {C.}},\ }\href@noop {} {\bibfield  {journal} {\bibinfo
  {journal} {Astron. \& Astrophys. Suppl. Ser.}\ }\textbf {\bibinfo {volume}
  {141}},\ \bibinfo {pages} {371} (\bibinfo {year} {2000})}\BibitemShut
  {NoStop}%
\bibitem [{\citenamefont {{Greene}}(1968)}]{1968AJS....73Q..15G}%
  \BibitemOpen
  \bibfield  {author} {\bibinfo {author} {\bibnamefont {{Greene}},
  \bibfnamefont {T.~F.}},\ }\href@noop {} {\bibfield  {journal} {\bibinfo
  {journal} {Astron. Journal Supp.}\ }\textbf {\bibinfo {volume} {73}},\
  \bibinfo {pages} {15} (\bibinfo {year} {1968})}\BibitemShut {NoStop}%
\bibitem [{\citenamefont {{Greenstein}}\ and\ \citenamefont
  {{Keenan}}(1958)}]{1958ApJ...127..172G}%
  \BibitemOpen
  \bibfield  {author} {\bibinfo {author} {\bibnamefont {{Greenstein}},
  \bibfnamefont {J.~L.}}\ and\ \bibinfo {author} {\bibnamefont {{Keenan}},
  \bibfnamefont {P.~C.}},\ }\href {\doibase 10.1086/146449} {\bibfield
  {journal} {\bibinfo  {journal} {Astrophys. Journal}\ }\textbf {\bibinfo
  {volume} {127}},\ \bibinfo {pages} {172} (\bibinfo {year}
  {1958})}\BibitemShut {NoStop}%
\bibitem [{\citenamefont {{Helfer}}\ and\ \citenamefont
  {{Wallerstein}}(1964)}]{1964ApJS....9...81H}%
  \BibitemOpen
  \bibfield  {author} {\bibinfo {author} {\bibnamefont {{Helfer}},
  \bibfnamefont {H.~L.}}\ and\ \bibinfo {author} {\bibnamefont {{Wallerstein}},
  \bibfnamefont {G.}},\ }\href {\doibase 10.1086/190098} {\bibfield  {journal}
  {\bibinfo  {journal} {Astrophys. J. Supp.}\ }\textbf {\bibinfo {volume}
  {9}},\ \bibinfo {pages} {81} (\bibinfo {year} {1964})}\BibitemShut {NoStop}%
\bibitem [{\citenamefont {{Helfer}}\ and\ \citenamefont
  {{Wallerstein}}(1968)}]{1968ApJS...16....1H}%
  \BibitemOpen
  \bibfield  {author} {\bibinfo {author} {\bibnamefont {{Helfer}},
  \bibfnamefont {H.~L.}}\ and\ \bibinfo {author} {\bibnamefont {{Wallerstein}},
  \bibfnamefont {G.}},\ }\href@noop {} {\bibfield  {journal} {\bibinfo
  {journal} {Astrophys. J. Suppl.}\ }\textbf {\bibinfo {volume} {16}},\
  \bibinfo {pages} {1} (\bibinfo {year} {1968})}\BibitemShut {NoStop}%
\bibitem [{\citenamefont {{Helfer}}, \citenamefont {{Wallerstein}},\ and\
  \citenamefont {{Greenstein}}(1959)}]{1959ApJ...129..700H}%
  \BibitemOpen
  \bibfield  {author} {\bibinfo {author} {\bibnamefont {{Helfer}},
  \bibfnamefont {H.~L.}}, \bibinfo {author} {\bibnamefont {{Wallerstein}},
  \bibfnamefont {G.}}, \ and\ \bibinfo {author} {\bibnamefont {{Greenstein}},
  \bibfnamefont {J.~L.}},\ }\href {\doibase 10.1086/146668} {\bibfield
  {journal} {\bibinfo  {journal} {\apj}\ }\textbf {\bibinfo {volume} {129}},\
  \bibinfo {pages} {700} (\bibinfo {year} {1959})}\BibitemShut {NoStop}%
\bibitem [{\citenamefont {{Jacobson}}, \citenamefont {{Friel}},\ and\
  \citenamefont {{Pilachowski}}(2011)}]{2011AJ....141...58J}%
  \BibitemOpen
  \bibfield  {author} {\bibinfo {author} {\bibnamefont {{Jacobson}},
  \bibfnamefont {H.~R.}}, \bibinfo {author} {\bibnamefont {{Friel}},
  \bibfnamefont {E.~D.}}, \ and\ \bibinfo {author} {\bibnamefont
  {{Pilachowski}}, \bibfnamefont {C.~A.}},\ }\href {\doibase
  10.1088/0004-6256/141/2/58} {\bibfield  {journal} {\bibinfo  {journal}
  {Astron. Journal}\ }\textbf {\bibinfo {volume} {141}},\ \bibinfo {eid} {58}
  (\bibinfo {year} {2011})}\BibitemShut {NoStop}%
\bibitem [{\citenamefont {{Korotin}}\ and\ \citenamefont
  {{Komarov}}(1989)}]{1989SvA....33..449K}%
  \BibitemOpen
  \bibfield  {author} {\bibinfo {author} {\bibnamefont {{Korotin}},
  \bibfnamefont {S.~A.}}\ and\ \bibinfo {author} {\bibnamefont {{Komarov}},
  \bibfnamefont {N.~S.}},\ }\href@noop {} {\bibfield  {journal} {\bibinfo
  {journal} {Soviet Astronomy}\ }\textbf {\bibinfo {volume} {33}},\ \bibinfo
  {pages} {449} (\bibinfo {year} {1989})}\BibitemShut {NoStop}%
\bibitem [{\citenamefont {{Kurucz}}(1993)}]{1993KurCD..13.....K}%
  \BibitemOpen
  \bibfield  {author} {\bibinfo {author} {\bibnamefont {{Kurucz}},
  \bibfnamefont {R.}},\ }\href@noop {} {\enquote {\bibinfo {title} {{ATLAS9
  Stellar Atmosphere Programs and 2 km/s grid.~Kurucz CD-ROM No.~13.~
  Cambridge, Mass.: Smithsonian Astrophysical Observatory, 1993.}}}\ }
  (\bibinfo {year} {1993})\BibitemShut {NoStop}%
\bibitem [{\citenamefont {{Lind}}\ \emph {et~al.}(2011)\citenamefont {{Lind}},
  \citenamefont {{Asplund}}, \citenamefont {{Barklem}},\ and\ \citenamefont
  {{Belyaev}}}]{2011A&A...528A.103L}%
  \BibitemOpen
  \bibfield  {author} {\bibinfo {author} {\bibnamefont {{Lind}}, \bibfnamefont
  {K.}}, \bibinfo {author} {\bibnamefont {{Asplund}}, \bibfnamefont {M.}},
  \bibinfo {author} {\bibnamefont {{Barklem}}, \bibfnamefont {P.~S.}}, \ and\
  \bibinfo {author} {\bibnamefont {{Belyaev}}, \bibfnamefont {A.~K.}},\ }\href
  {\doibase 10.1051/0004-6361/201016095} {\bibfield  {journal} {\bibinfo
  {journal} {Astron. and Astrophys.}\ }\textbf {\bibinfo {volume} {528}},\
  \bibinfo {pages} {A103+} (\bibinfo {year} {2011})},\ \Eprint
  {http://arxiv.org/abs/1102.2160} {arXiv:1102.2160 [astro-ph.SR]} \BibitemShut
  {NoStop}%
\bibitem [{\citenamefont {{Mashonkina}}, \citenamefont {{Shimanski{\u i}}},\
  and\ \citenamefont {{Sakhibullin}}(2000)}]{2000ARep...44..790M}%
  \BibitemOpen
  \bibfield  {author} {\bibinfo {author} {\bibnamefont {{Mashonkina}},
  \bibfnamefont {L.~I.}}, \bibinfo {author} {\bibnamefont {{Shimanski{\u i}}},
  \bibfnamefont {V.~V.}}, \ and\ \bibinfo {author} {\bibnamefont
  {{Sakhibullin}}, \bibfnamefont {N.~A.}},\ }\href {\doibase 10.1134/1.1327637}
  {\bibfield  {journal} {\bibinfo  {journal} {Astronomy Reports}\ }\textbf
  {\bibinfo {volume} {44}},\ \bibinfo {pages} {790} (\bibinfo {year}
  {2000})}\BibitemShut {NoStop}%
\bibitem [{\citenamefont {{Maurice}}\ \emph {et~al.}(1987)\citenamefont
  {{Maurice}}, \citenamefont {{Andersen}}, \citenamefont {{Ardeberg}},
  \citenamefont {{Bardin}}, \citenamefont {{Imbert}}, \citenamefont
  {{Lindgren}}, \citenamefont {{Martin}}, \citenamefont {{Mayor}},
  \citenamefont {{Nordstrom}}, \citenamefont {{Prevot}}, \citenamefont
  {{Rebeirot}},\ and\ \citenamefont {{Rousseau}}}]{1987A&AS...67..423M}%
  \BibitemOpen
  \bibfield  {author} {\bibinfo {author} {\bibnamefont {{Maurice}},
  \bibfnamefont {E.}}, \bibinfo {author} {\bibnamefont {{Andersen}},
  \bibfnamefont {J.}}, \bibinfo {author} {\bibnamefont {{Ardeberg}},
  \bibfnamefont {A.}}, \bibinfo {author} {\bibnamefont {{Bardin}},
  \bibfnamefont {C.}}, \bibinfo {author} {\bibnamefont {{Imbert}},
  \bibfnamefont {M.}}, \bibinfo {author} {\bibnamefont {{Lindgren}},
  \bibfnamefont {H.}}, \bibinfo {author} {\bibnamefont {{Martin}},
  \bibfnamefont {M.}}, \bibinfo {author} {\bibnamefont {{Mayor}}, \bibfnamefont
  {M.}}, \bibinfo {author} {\bibnamefont {{Nordstrom}}, \bibfnamefont {B.}},
  \bibinfo {author} {\bibnamefont {{Prevot}}, \bibfnamefont {L.}}, \bibinfo
  {author} {\bibnamefont {{Rebeirot}}, \bibfnamefont {E.}}, \ and\ \bibinfo
  {author} {\bibnamefont {{Rousseau}}, \bibfnamefont {J.}},\ }\href@noop {}
  {\bibfield  {journal} {\bibinfo  {journal} {Astron. \& Astrophys. Suppl.
  Ser.}\ }\textbf {\bibinfo {volume} {67}},\ \bibinfo {pages} {423} (\bibinfo
  {year} {1987})}\BibitemShut {NoStop}%
\bibitem [{\citenamefont {{Mishenina}}\ \emph {et~al.}(2004)\citenamefont
  {{Mishenina}}, \citenamefont {{Soubiran}}, \citenamefont {{Kovtyukh}},\ and\
  \citenamefont {{Korotin}}}]{2004A&A...418..551M}%
  \BibitemOpen
  \bibfield  {author} {\bibinfo {author} {\bibnamefont {{Mishenina}},
  \bibfnamefont {T.~V.}}, \bibinfo {author} {\bibnamefont {{Soubiran}},
  \bibfnamefont {C.}}, \bibinfo {author} {\bibnamefont {{Kovtyukh}},
  \bibfnamefont {V.~V.}}, \ and\ \bibinfo {author} {\bibnamefont {{Korotin}},
  \bibfnamefont {S.~A.}},\ }\href {\doibase 10.1051/0004-6361:20034454}
  {\bibfield  {journal} {\bibinfo  {journal} {Astron. \& Astrophys.}\ }\textbf
  {\bibinfo {volume} {418}},\ \bibinfo {pages} {551} (\bibinfo {year}
  {2004})}\BibitemShut {NoStop}%
\bibitem [{\citenamefont {{Pakhomov}}(2012)}]{2012AstL...38..101P}%
  \BibitemOpen
  \bibfield  {author} {\bibinfo {author} {\bibnamefont {{Pakhomov}},
  \bibfnamefont {Y.~V.}},\ }\href {\doibase 10.1134/S1063773712020053}
  {\bibfield  {journal} {\bibinfo  {journal} {Astronomy Letters}\ }\textbf
  {\bibinfo {volume} {38}},\ \bibinfo {pages} {101} (\bibinfo {year}
  {2012})}\BibitemShut {NoStop}%
\bibitem [{\citenamefont {{Pakhomov}}, \citenamefont {{Antipova}},\ and\
  \citenamefont {{Boyarchuk}}(2011)}]{2011ARep...55..256P}%
  \BibitemOpen
  \bibfield  {author} {\bibinfo {author} {\bibnamefont {{Pakhomov}},
  \bibfnamefont {Y.~V.}}, \bibinfo {author} {\bibnamefont {{Antipova}},
  \bibfnamefont {L.~I.}}, \ and\ \bibinfo {author} {\bibnamefont {{Boyarchuk}},
  \bibfnamefont {A.~A.}},\ }\href {\doibase 10.1134/S106377291103005X}
  {\bibfield  {journal} {\bibinfo  {journal} {Astronomy Reports}\ }\textbf
  {\bibinfo {volume} {55}},\ \bibinfo {pages} {256} (\bibinfo {year}
  {2011})}\BibitemShut {NoStop}%
\bibitem [{\citenamefont {{Pakhomov}}\ \emph
  {et~al.}(2009{\natexlab{a}})\citenamefont {{Pakhomov}}, \citenamefont
  {{Antipova}}, \citenamefont {{Boyarchuk}}, \citenamefont {{Bizyaev}},
  \citenamefont {{Zhao}},\ and\ \citenamefont {{Liang}}}]{2009ARep...53..660P}%
  \BibitemOpen
  \bibfield  {author} {\bibinfo {author} {\bibnamefont {{Pakhomov}},
  \bibfnamefont {Y.~V.}}, \bibinfo {author} {\bibnamefont {{Antipova}},
  \bibfnamefont {L.~I.}}, \bibinfo {author} {\bibnamefont {{Boyarchuk}},
  \bibfnamefont {A.~A.}}, \bibinfo {author} {\bibnamefont {{Bizyaev}},
  \bibfnamefont {D.~V.}}, \bibinfo {author} {\bibnamefont {{Zhao}},
  \bibfnamefont {G.}}, \ and\ \bibinfo {author} {\bibnamefont {{Liang}},
  \bibfnamefont {Y.}},\ }\href {\doibase 10.1134/S1063772909070087} {\bibfield
  {journal} {\bibinfo  {journal} {Astronomy Reports}\ }\textbf {\bibinfo
  {volume} {53}},\ \bibinfo {pages} {660} (\bibinfo {year}
  {2009}{\natexlab{a}})}\BibitemShut {NoStop}%
\bibitem [{\citenamefont {{Pakhomov}}\ \emph
  {et~al.}(2009{\natexlab{b}})\citenamefont {{Pakhomov}}, \citenamefont
  {{Antipova}}, \citenamefont {{Boyarchuk}}, \citenamefont {{Zhao}},\ and\
  \citenamefont {{Liang}}}]{2009ARep...53..685P}%
  \BibitemOpen
  \bibfield  {author} {\bibinfo {author} {\bibnamefont {{Pakhomov}},
  \bibfnamefont {Y.~V.}}, \bibinfo {author} {\bibnamefont {{Antipova}},
  \bibfnamefont {L.~I.}}, \bibinfo {author} {\bibnamefont {{Boyarchuk}},
  \bibfnamefont {A.~A.}}, \bibinfo {author} {\bibnamefont {{Zhao}},
  \bibfnamefont {G.}}, \ and\ \bibinfo {author} {\bibnamefont {{Liang}},
  \bibfnamefont {Y.}},\ }\href {\doibase 10.1134/S1063772909080010} {\bibfield
  {journal} {\bibinfo  {journal} {Astronomy Reports}\ }\textbf {\bibinfo
  {volume} {53}},\ \bibinfo {pages} {685} (\bibinfo {year}
  {2009}{\natexlab{b}})}\BibitemShut {NoStop}%
\bibitem [{\citenamefont {{Robin}}\ \emph {et~al.}(1996)\citenamefont
  {{Robin}}, \citenamefont {{Haywood}}, \citenamefont {{Creze}}, \citenamefont
  {{Ojha}},\ and\ \citenamefont {{Bienayme}}}]{1996A&A...305..125R}%
  \BibitemOpen
  \bibfield  {author} {\bibinfo {author} {\bibnamefont {{Robin}}, \bibfnamefont
  {A.~C.}}, \bibinfo {author} {\bibnamefont {{Haywood}}, \bibfnamefont {M.}},
  \bibinfo {author} {\bibnamefont {{Creze}}, \bibfnamefont {M.}}, \bibinfo
  {author} {\bibnamefont {{Ojha}}, \bibfnamefont {D.~K.}}, \ and\ \bibinfo
  {author} {\bibnamefont {{Bienayme}}, \bibfnamefont {O.}},\ }\href@noop {}
  {\bibfield  {journal} {\bibinfo  {journal} {Astron. \& Astrophys.}\ }\textbf
  {\bibinfo {volume} {305}},\ \bibinfo {pages} {125} (\bibinfo {year}
  {1996})}\BibitemShut {NoStop}%
\bibitem [{\citenamefont {{Schopp}}(1954)}]{1954AJ.....59R.192S}%
  \BibitemOpen
  \bibfield  {author} {\bibinfo {author} {\bibnamefont {{Schopp}},
  \bibfnamefont {J.~D.}},\ }\href {\doibase 10.1086/107079} {\bibfield
  {journal} {\bibinfo  {journal} {Astron. Journal}\ }\textbf {\bibinfo {volume}
  {59}},\ \bibinfo {pages} {192} (\bibinfo {year} {1954})}\BibitemShut
  {NoStop}%
\bibitem [{\citenamefont {{Smiljanic}}(2012)}]{2012MNRAS.422.1562S}%
  \BibitemOpen
  \bibfield  {author} {\bibinfo {author} {\bibnamefont {{Smiljanic}},
  \bibfnamefont {R.}},\ }\href {\doibase 10.1111/j.1365-2966.2012.20729.x}
  {\bibfield  {journal} {\bibinfo  {journal} {Monthly Notices Roy. Astron.
  Soc.}\ }\textbf {\bibinfo {volume} {422}},\ \bibinfo {pages} {1562} (\bibinfo
  {year} {2012})},\ \Eprint {http://arxiv.org/abs/1202.2200} {arXiv:1202.2200
  [astro-ph.SR]} \BibitemShut {NoStop}%
\bibitem [{\citenamefont {{Takeda}}\ and\ \citenamefont
  {{Takada-Hidai}}(1994)}]{1994PASJ...46..395T}%
  \BibitemOpen
  \bibfield  {author} {\bibinfo {author} {\bibnamefont {{Takeda}},
  \bibfnamefont {Y.}}\ and\ \bibinfo {author} {\bibnamefont {{Takada-Hidai}},
  \bibfnamefont {M.}},\ }\href@noop {} {\bibfield  {journal} {\bibinfo
  {journal} {Pub. Astron. Soc. of Japan}\ }\textbf {\bibinfo {volume} {46}},\
  \bibinfo {pages} {395} (\bibinfo {year} {1994})}\BibitemShut {NoStop}%
\bibitem [{\citenamefont {{van Leeuwen}}(2007)}]{2007A&A...474..653V}%
  \BibitemOpen
  \bibfield  {author} {\bibinfo {author} {\bibnamefont {{van Leeuwen}},
  \bibfnamefont {F.}},\ }\href {\doibase 10.1051/0004-6361:20078357} {\bibfield
   {journal} {\bibinfo  {journal} {Astron. \& Astrophys.}\ }\textbf {\bibinfo
  {volume} {474}},\ \bibinfo {pages} {653} (\bibinfo {year}
  {2007})}\BibitemShut {NoStop}%
\bibitem [{\citenamefont {Wallerstein}\ \emph {et~al.}(1997)\citenamefont
  {Wallerstein}, \citenamefont {Iben}, \citenamefont {Parker}, \citenamefont
  {Boesgaard}, \citenamefont {Hale}, \citenamefont {Champagne}, \citenamefont
  {Barnes}, \citenamefont {K\"appeler}, \citenamefont {Smith}, \citenamefont
  {Hoffman}, \citenamefont {Timmes}, \citenamefont {Sneden}, \citenamefont
  {Boyd}, \citenamefont {Meyer},\ and\ \citenamefont
  {Lambert}}]{RevModPhys..69.995W}%
  \BibitemOpen
  \bibfield  {author} {\bibinfo {author} {\bibnamefont {Wallerstein},
  \bibfnamefont {G.}}, \bibinfo {author} {\bibnamefont {Iben}, \bibfnamefont
  {I.}}, \bibinfo {author} {\bibnamefont {Parker}, \bibfnamefont {P.}},
  \bibinfo {author} {\bibnamefont {Boesgaard}, \bibfnamefont {A.~M.}}, \bibinfo
  {author} {\bibnamefont {Hale}, \bibfnamefont {G.~M.}}, \bibinfo {author}
  {\bibnamefont {Champagne}, \bibfnamefont {A.~E.}}, \bibinfo {author}
  {\bibnamefont {Barnes}, \bibfnamefont {C.~A.}}, \bibinfo {author}
  {\bibnamefont {K\"appeler}, \bibfnamefont {F.}}, \bibinfo {author}
  {\bibnamefont {Smith}, \bibfnamefont {V.~V.}}, \bibinfo {author}
  {\bibnamefont {Hoffman}, \bibfnamefont {R.~D.}}, \bibinfo {author}
  {\bibnamefont {Timmes}, \bibfnamefont {F.~X.}}, \bibinfo {author}
  {\bibnamefont {Sneden}, \bibfnamefont {C.}}, \bibinfo {author} {\bibnamefont
  {Boyd}, \bibfnamefont {R.~N.}}, \bibinfo {author} {\bibnamefont {Meyer},
  \bibfnamefont {B.~S.}}, \ and\ \bibinfo {author} {\bibnamefont {Lambert},
  \bibfnamefont {D.~L.}},\ }\href {\doibase 10.1103/RevModPhys.69.995}
  {\bibfield  {journal} {\bibinfo  {journal} {Rev. Mod. Phys.}\ }\textbf
  {\bibinfo {volume} {69}},\ \bibinfo {pages} {995} (\bibinfo {year}
  {1997})}\BibitemShut {NoStop}%
\bibitem [{\citenamefont {{Weiss}}\ and\ \citenamefont
  {{Charbonnel}}(2004)}]{2004MmSAI..75..347W}%
  \BibitemOpen
  \bibfield  {author} {\bibinfo {author} {\bibnamefont {{Weiss}}, \bibfnamefont
  {A.}}\ and\ \bibinfo {author} {\bibnamefont {{Charbonnel}}, \bibfnamefont
  {C.}},\ }\href@noop {} {\bibfield  {journal} {\bibinfo  {journal} {Mem del.
  Soc. Astron. Ital.}\ }\textbf {\bibinfo {volume} {75}},\ \bibinfo {pages}
  {347} (\bibinfo {year} {2004})}\BibitemShut {NoStop}%
\end{thebibliography}%

\end{document}